\documentclass[aps,prx,nofootinbib,notitlepage,superscriptaddress]{revtex4-1}

\usepackage{graphicx}
\usepackage{bm}
\usepackage{amsmath}
\usepackage{physics}
\usepackage{amssymb}
\usepackage{amsfonts}
\usepackage{amsthm}
\usepackage{bbm}
\usepackage{hyperref}
\usepackage{braket}
\usepackage[normalem]{ulem}
\usepackage{wrapfig}
\usepackage{tikz}
\usepackage{dsfont}
\usepackage{comment}
\usepackage{thmtools,thm-restate}
\usepackage[export]{adjustbox}
\usepackage{mathtools}
\usepackage[noend]{algcompatible}
\usepackage{algorithm}
\usepackage{pdfpages}
\usepackage{lineno}
\usepackage{outlines}

\definecolor{darkblue}{rgb}{0,0,0.5}
\hypersetup{
colorlinks=true,
linkcolor=black,
filecolor=blue,
citecolor=darkblue,  
urlcolor=black,
}

\usepackage[strict]{changepage}

\usepackage{setspace}
\let\Algorithm\algorithm
\renewcommand\algorithm[1][]{\Algorithm[#1]\setstretch{1.25}}

\newtheorem{theorem}{Theorem}
\newtheorem{extension}{Extension}
\newtheorem{definition}{Definition}
\newtheorem{corollary}{Corollary}
\newtheorem{lemma}{Lemma}

\newtheorem*{theorem*}{Theorem}

\newtheorem*{task*}{Task}
\newtheorem*{proposition*}{Proposition}

\DeclarePairedDelimiter{\bbra}{\langle \! \langle}{\rVert}
\DeclarePairedDelimiter{\kket}{\lVert}{\rangle \! \rangle}
\newcommand{\iipp}[2]{\langle \! \langle #1 \lVert #2 \rangle \! \rangle}
\newcommand{\be}{\begin{equation}}
\newcommand{\ee}{\end{equation}}

\newcommand{\lr}[1]{\left( #1\right)}
\newcommand{\OO}{O}
\newcommand{\cU}{\mathcal{U}}
\newcommand{\bP}{\mathcal{P}}

\newcommand{\keto}[1]{\lVert #1 \rangle \! \rangle}

\newcommand{\poly}[1]{\emph{poly}\lr{#1} }

\makeatletter

\newenvironment{subtheorem}[1]{%
  \def\subtheoremcounter{#1}%
  \refstepcounter{#1}%
  \protected@edef\theparentnumber{\csname the#1\endcsname}%
  \setcounter{parentnumber}{\value{#1}}%
  \setcounter{#1}{0}%
  \expandafter\def\csname the#1\endcsname{\theparentnumber\alph{#1}}%
  \ignorespaces
}{%
  \setcounter{\subtheoremcounter}{\value{parentnumber}}%
  \ignorespacesafterend
}
\makeatother
\newcounter{parentnumber}

\newcommand{\FigNoisyCircuit}
{
\begin{figure}
\centering
\includegraphics[width=\textwidth]{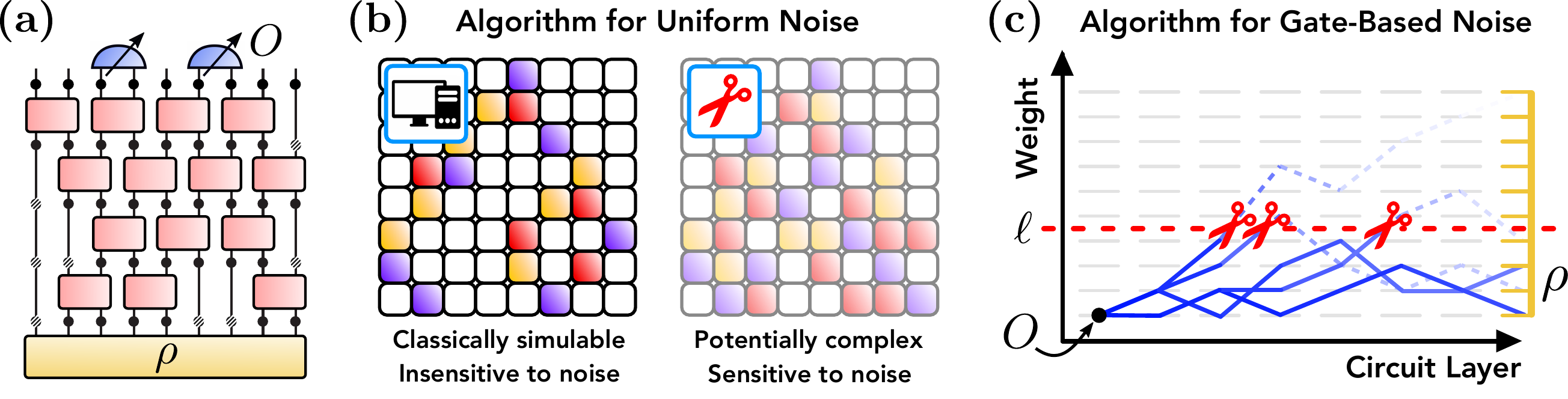}
\caption{\textbf{(a)} Schematic of a noisy quantum circuit.
The input state $\rho$ is acted on by a circuit of arbitrary two-qubit gates (pink) and local noise channels (dots), and concludes with measurement of an observable $O$. 
In the gate-based noise model, the noise only acts on a qubit when a gate is performed (black dots).
In the uniform noise model, even when the qubit is idle, noise can occur (dashed dots). 
\textbf{(b)} For circuits with uniform noise, our classical algorithm decomposes the expectation value of $O$ as a sum of Pauli paths.
We depict each path by a space-time grid, where each square denotes whether the path is the identity (white) or $X$, $Y$, or $Z$ (red, yellow, or purple) at that qubit and circuit layer.
Our algorithm computes the sum of all low-weight paths (left), and truncates high-weight paths (right) since they are strongly damped by noise (fading).
\textbf{(c)} For circuits with gate-based noise, our algorithm instead simulates the Heisenberg time-evolution of $O$ within the subspace of low-weight Pauli operators.
That is, at each layer, we truncate all Pauli operators with weight above a threshold $\ell$ (dashed red line).
The plot depicts the weight of various components of the time-evolved operator (blue lines), as they are acted on by noise (fading) and, potentially, truncated by our algorithm (red scissors).
}
\label{fig: noisy circuit}
\end{figure}
}

\newcommand{\FigProof}
{
\begin{figure}
\centering
\includegraphics[width=\textwidth]{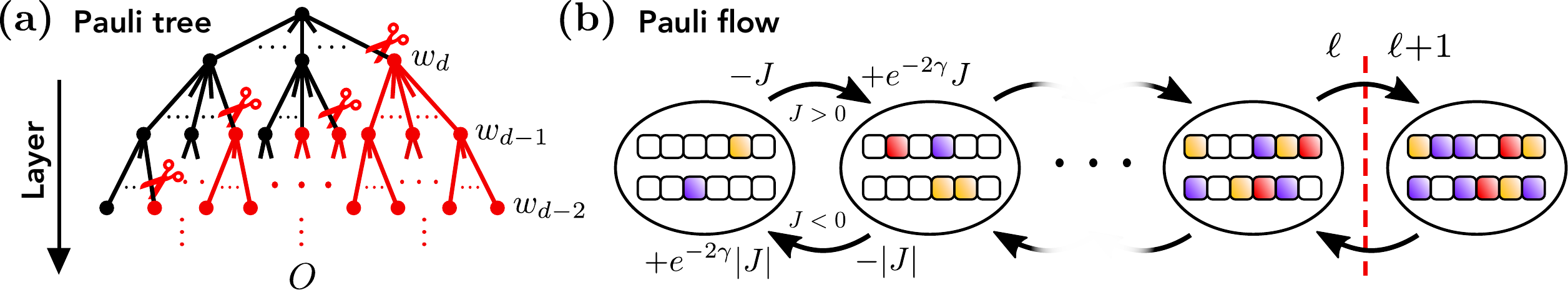}
\caption{\textbf{(a)} The Pauli tree framework used to prove Theorem~\ref{thm 1}, which groups Pauli paths according to their weight $w_t$ at each circuit layer $t$.
Our algorithm truncates paths with a summed weight above $\ell$.
We group these truncations into ${\ell \choose d}$ individual truncations (red scissors), which enables a tight bound on the algorithm's error. In the example shown, the  branchings at each node correspond to weights $w = 1,\ldots,5$ and we set $\ell = 6, d = 2$.
\textbf{(b)} To prove Theorem~\ref{thm 2}, we analyze the flow of Pauli operators from one weight to another under circuit gates.
Here, each bubble depicts the set of Pauli operators of a given weight, and arrows indicate flow from one weight to another.
We show that this flow is lossy in the presence of gate-based noise, in the sense that a decrease  $-J$ in support at weight $w$ can increase the support at weight $w+1$ by at most $e^{-2\gamma} J$. 
This leads to our bound on the cumulative operator weight distribution in Lemma~\ref{lemma: operator norm noise}.
}
\label{fig: proof}
\end{figure}
}

\newcommand{\FigErrorMitigation}
{
\begin{figure}
\centering
\includegraphics[width=\textwidth]{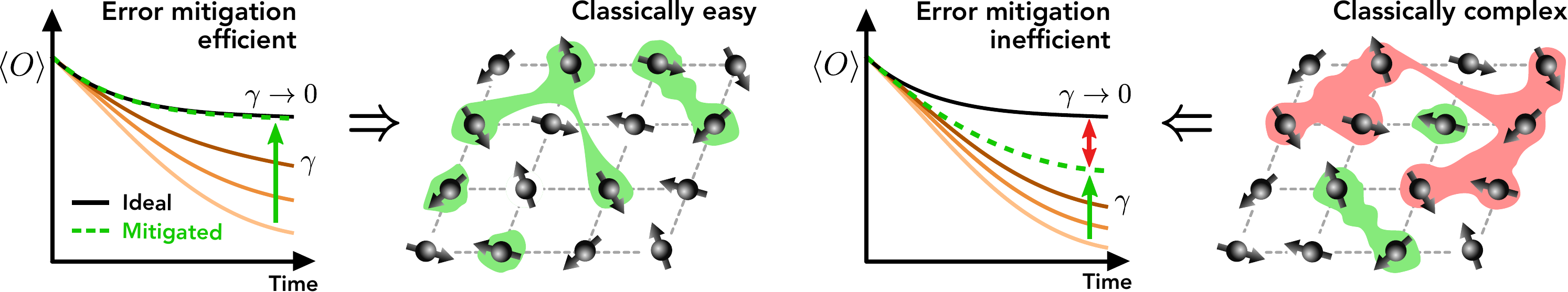}
\caption{Schematic of our results' implications for quantum error mitigation.
For concreteness, we depict a specific error mitigation strategy, zero noise extrapolation, in which one measures an expectation value $\langle O \rangle$ for several different noise rates (orange lines), and performs an extrapolation (green arrow and dashed line) to estimate the ideal expectation value (black line).
\textbf{(a)} If error mitigation succeeds in recovering the ideal expectation value, then the expectation value must be dominated by low-weight Pauli operators (green).
Thus, our classical algorithm can also compute the ideal expectation value.
\textbf{(b)} On the other hand, if the ideal expectation value is hard to compute classically, then it must contain contributions from high-weight Pauli operators (red).
Error mitigation cannot capture these contributions, since they are exponentially suppressed by noise.
Thus, the extrapolated expectation value necessarily differs from the ideal value (red arrows).
}
\label{fig: error mitigation}
\end{figure}
}

\makeatletter
\AtBeginDocument{\let\LS@rot\@undefined}
\makeatother

\allowdisplaybreaks

\begin{document}

\title{A polynomial-time classical algorithm for noisy quantum circuits}

\author{Thomas Schuster}
\thanks{These authors contributed equally to this work.}
\affiliation{Walter Burke Institute for Theoretical Physics and Institute for Quantum Information and Matter, California Institute of Technology,
Pasadena, California 91125, USA}

\author{Chao Yin}
\thanks{These authors contributed equally to this work.}
\affiliation{Department of Physics and Center for Theory of Quantum Matter,
University of Colorado, Boulder, CO 80309, USA}

\author{Xun Gao}
\affiliation{Department of Physics and Center for Theory of Quantum Matter,
University of Colorado, Boulder, CO 80309, USA}
\affiliation{JILA,
University of Colorado, Boulder, CO 80309, USA}

\author{Norman Y. Yao}
\affiliation{Department of Physics, Harvard University, Cambridge, MA 02138, USA}

\date{\today}

\begin{abstract}
We provide a polynomial-time classical algorithm for  noisy quantum circuits.
The algorithm computes the expectation value of any observable for any circuit, with a small average error over input states drawn from an ensemble (e.g.~the computational basis).
Our approach is based upon the intuition that noise exponentially damps non-local correlations relative to local correlations. 
This enables one to classically simulate a noisy quantum circuit by only keeping track of the dynamics of local quantum information.
Our algorithm also enables sampling from the output distribution of a circuit in quasi-polynomial time, so long as the distribution anti-concentrates.
A number of practical implications are discussed, including a fundamental limit on the efficacy of noise mitigation strategies: for constant noise rates, any quantum circuit for which error mitigation is efficient on most input states, is also classically simulable on most input states. 
\end{abstract}

\maketitle


\section{Introduction}

Quantum computers are believed to yield exponential computational advantages over their classical counterparts for certain tasks~\cite{feynman2018simulating,shor1994algorithms,bauer2020quantum,dalzell2023quantum}.
However, near-term quantum devices are inevitably impacted by noise.
This raises a key question~\cite{aharonov1996polynomial,aharonov2000quantum,harrow2003robustness,kempe2008upper,shor1996fault,aharonov1997fault,kitaev1997quantum,knill1998resilient,aharonov1996limitations,ben2013quantum,dalzell2021random,deshpande2022tight,chen2022complexity,huang2020classical,liu2021closing,pan2021simulating,villalonga2020establishing,cheng2021simulating,noh2020efficient,chen2023optimized,cheng2023efficient,bremner2017achieving,rajakumar2024polynomial,gao2018efficient,barak2020spoofing,gao2021limitations,aharonov2023polynomial,yung2017can,fontana2023classical,shao2023simulating,bouland2022noise,mele2024noise,movassagh2021theory,bouland2022noise,fujii2016noise,kalai2014gaussian,garcia2019simulating,qi2020regimes,oh2023classical,oh2023tensor,hangleiter2023bell,tanggara2024classically,ermakov2024unified,kuprov2007polynomially,karabanov2011accuracy,rakovszky2022dissipation,yoo2023open,vovk2022entanglement}: To what degree does noise \emph{fundamentally} limit any quantum advantage over classical computation?

The answer to this question is well understood in two limits. 
On the one hand, for noise rates above a large threshold, seminal works have shown that noisy quantum circuits can be simulated classically~\cite{aharonov1996polynomial,aharonov2000quantum,harrow2003robustness,kempe2008upper}.
On the other hand, for sufficiently low noise rates,  one can utilize quantum error correction to perform fault-tolerant computation~\cite{shor1996fault,aharonov1997fault,kitaev1997quantum,knill1998resilient}.
Nevertheless, the reality of modern quantum experiments lies precisely in between these two limits:  Noise rates are small but error correction is  not typically employed.
Progress in this intermediate regime has been restricted to several relatively specific contexts:  bounds on the  circuit depth but not the complexity~\cite{aharonov1996limitations,ben2013quantum,dalzell2021random,deshpande2022tight}, tasks that assume  access to a noiseless oracle~\cite{chen2022complexity}, hardness results for sampling with extremely high precision~\cite{bouland2022noise,dalzell2024random} or error detection~\cite{fujii2016noise,bremner2017achieving,paletta2024robust}, and classical algorithms for instantaneous quantum polynomial (IQP) circuits~\cite{bremner2017achieving,rajakumar2024polynomial} and  random quantum circuits~\cite{gao2018efficient,barak2020spoofing,gao2021limitations,aharonov2023polynomial,yung2017can,fontana2023classical,shao2023simulating,bouland2022noise,mele2024noise}.
To date, however, there are few rigorous results on the computational power of general quantum circuits with low noise rates but without error correction.


In this work, we provide a  classical algorithm for computing expectation values in \emph{any} noisy quantum circuit on \emph{most} input states.
The restriction to ``most'' input states is fundamental, since for a fixed input state and a sufficiently low noise rate, one can immediately perform quantum error correction.
To this end, with a high  probability over input states drawn from an ensemble, our algorithm succeeds in computing the  expectation value of any observable (to within small error) for any noisy quantum circuit.
We emphasize that our restriction to ensembles of input states is relatively weak---a wide range of ensembles are allowed, including the computational basis states.
The algorithm runs in either polynomial or quasi-polynomial time, depending on the observable and  noise model of interest.
%

To compute the expectation value, our algorithm simulates the Heisenberg time-evolution of the observable in the Pauli basis.
Building upon seminal recent algorithms for random quantum circuits~\cite{gao2018efficient,barak2020spoofing,gao2021limitations,aharonov2023polynomial,yung2017can,fontana2023classical,shao2023simulating,bouland2022noise,mele2024noise} and many-body dynamics~\cite{kuprov2007polynomially,karabanov2011accuracy,white2018quantum,vovk2022entanglement,ye2020emergent,rakovszky2022dissipation,von2022operator,white2023effective,yoo2023open,klein2022time,artiaco2024efficient,ermakov2024unified}, the key insight underlying our algorithm is the close connection between sets of Pauli operators that are hard-to-classically-simulate, and those that are strongly affected by noise. 
By carefully truncating such high-weight Pauli operators, we achieve a provably efficient and accurate simulation.
We emphasize that, despite the prevalence of Pauli truncation methods for nearly six years to date~\cite{gao2018efficient}, our work provides the first rigorous proof that such methods can succeed for non-random circuits.
This is enabled by substantially new proof techniques compared to any previous work.

Our results have wide-ranging consequences for quantum experiments.
First, we show that, for noise rates that do not improve as the number of qubits increases, any strategy for quantum error mitigation~\cite{temme2017error,li2017efficient,cai2023quantum} can only scale efficiently~\cite{quek2022exponentially,takagi2022fundamental,takagi2023universal,tsubouchi2023universal} for quantum circuits that are easy to classically simulate. 
Like our main results, this statement holds for any quantum circuit on most input states.
Second, we build upon our results to construct a simple and strict test for whether a given quantum circuit can exhibit quantum advantage.
We find that any circuit exhibiting advantage on most input states must be macroscopically sensitive to noise.
Finally, we provide extensions of our algorithm to sampling from circuits with output distributions that anti-concentrate, as well as to computing expectation values in random quantum circuits with non-unital noise~\cite{fefferman2023effect,mele2024noise}, which improves in both runtime and generality compared to existing algorithms~\cite{mele2024noise}.

\FigNoisyCircuit

\vspace{5mm}
{\renewcommand\addcontentsline[3]{} \section{Classical algorithm for  noisy quantum circuits} \label{sec: exp val}}

Consider a noisy quantum circuit on $n$ qubits, of the form
\begin{equation}
	\mathcal{C} \{ \rho \} =  \mathcal{D}_0 \{ U_1  \, \mathcal{D}_1 \{ \ldots  U_d \,  \mathcal{D}_d \{ \rho \}  \, U_d^\dagger  \ldots \}  \, U_1^\dagger \},
\end{equation}
where $\rho$ is the input state, $U_t$ are depth-1 unitaries of arbitrary two-qubit gates, and $\mathcal{D}_t$ are local depolarizing noise channels.
Here, $d$ is the circuit depth, and $t =1,\ldots,d$ indexes the circuit layers in reverse order.
After the circuit is applied, we are interested in either computing the expectation value of an observable $O$, $\tr( \mathcal{C} \{ \rho \} O )$, or sampling from the outcomes when $\mathcal{C}\{\rho\}$ is measured in the computational basis.

We consider two archetypal noise models [Fig.~\ref{fig: noisy circuit}(a)]: uniform noise and gate-based noise. 
In the \emph{uniform noise} case, every qubit is affected by noise at every circuit layer.
That is, we take $\mathcal{D}_t  = \bigotimes_{j = 1}^n \mathcal{D}_{j} $, where $\mathcal{D}_j\{ \rho \} = e^{-\gamma} \rho + (1-e^{-\gamma}) \tr_j(\rho)$ is a local depolarizing channel of strength $\gamma$.
This noise model is relevant for many current experiments, in which idle errors affect qubits even when they are not involved in gates.
However, the uniform noise model only allows computation up to a depth logarithmic in the number of qubits, after which the system is close to the maximally mixed state~\cite{aharonov1996limitations,fn2}.
To this end, we also consider the so-called \emph{gate-based noise} model, in which qubits are only affected by noise when they participate in non-identity gates, $\mathcal{D}_t  = \bigotimes_{j \in U_t} \mathcal{D}_{j} $.
This model allows for ``fresh qubits'' unaffected by idle errors to be inserted at any circuit layer, which is the standard setting for proofs of fault-tolerant quantum computation~\cite{shor1996fault,aharonov1997fault,kitaev1997quantum,knill1998resilient,fn5}.
%
In both noise models, we incorporate read-out noise by taking the final noise channel $\mathcal{D}_0$ to act on all qubits.

\begin{table}[t] \label{table}
\vspace{-2.3mm}
\centering 
\def\arraystretch{1.1}
\setlength\tabcolsep{1.4mm}
\begin{tabular}{| c || c | c | c | c | c | c |} 
\hline 
& \def\arraystretch{0.2} \begin{tabular}{@{}c@{}}IQP \\ circuits \end{tabular} 
& \def\arraystretch{0.3} \begin{tabular}{@{}c@{}}Random \\ circuits \end{tabular} 
& \def\arraystretch{0.65} \begin{tabular}{@{}c@{}} Any circuit, \\ most inputs \\ (uniform noise) \end{tabular} 
& \def\arraystretch{0.65} \begin{tabular}{@{}c@{}} Any circuit, \\ most inputs \\ (gate noise) \end{tabular} 
& \def\arraystretch{0.65} \begin{tabular}{@{}c@{}} Any circuit, \\ any input \\ (uniform noise) \end{tabular} 
& \def\arraystretch{0.65} \begin{tabular}{@{}c@{}} Any circuit, \\ \, any input \, \\ (gate noise) \end{tabular} \rule{0pt}{8.15mm} \\ [5.9mm]
\hline \hline
\begin{tabular}{@{}c@{}} Expectation \\ values \end{tabular} 
& \def\arraystretch{0.75} \begin{tabular}{@{}c@{}} poly $n$ \\ Ref.~\cite{bremner2017achieving}  \end{tabular} 
& \def\arraystretch{0.75}\begin{tabular}{@{}c@{}} poly $n$ \\ Refs.~\cite{gao2018efficient,aharonov2023polynomial}   \end{tabular} 
& \def\arraystretch{0.75} \begin{tabular}{@{}c@{}} poly $n$ \\ Theorem~\ref{thm 1}   \end{tabular} 
& \begin{tabular}{@{}c@{}} quasi-poly $n$  \\ Theorem~\ref{thm 2} \end{tabular}
& \begin{tabular}{@{}c@{}} $\widetilde{\text{QNC}}_1$ \\ Ref.~\cite{aharonov1996limitations} \end{tabular} 
& \begin{tabular}{@{}c@{}} BQP \\ Refs.~\cite{shor1996fault,aharonov1997fault,kitaev1997quantum,knill1998resilient} \end{tabular} \rule{0pt}{8mm} \\ [5.1mm]
\hline
\def\arraystretch{0.75} \begin{tabular}{@{}c@{}} Sampling \\ ($^*$: with anti- \\ concentration)  \end{tabular} 
& \begin{tabular}{@{}c@{}} quasi-poly $n^*$ \\ Ref.~\cite{bremner2017achieving} \end{tabular} 
& \begin{tabular}{@{}c@{}} poly $n^*$ \\ Ref.~\cite{aharonov2023polynomial} \end{tabular} 
& \begin{tabular}{@{}c@{}}  quasi-poly $n^*$ \\ Theorem~\ref{thm sampling a} \end{tabular} 
& \begin{tabular}{@{}c@{}}  quasi-poly $n^*$ \\ Theorem~\ref{thm sampling a} \end{tabular} 
& \begin{tabular}{@{}c@{}} Samp$\widetilde{\text{QNC}}_1$ \\ Ref.~\cite{aharonov1996limitations} \end{tabular}
& \begin{tabular}{@{}c@{}} SampBQP \\ Refs.~\cite{shor1996fault,aharonov1997fault,kitaev1997quantum,knill1998resilient} \end{tabular} \rule{0pt}{8.35mm} \\[5.5mm]
\hline 
\end{tabular}
\caption{Known results on the computational complexity of various classes of noisy quantum circuits, taking the standard limits $d,\varepsilon^{-1} = \text{poly}(n)$ and $\gamma^{-1} = \mathcal{O}(1)$. 
In the first three columns, expectation values are with respect to observables that are sums of polynomially many Pauli operators; each runtime increases to quasi-polynomial otherwise.
In the final two columns, entries denote hardness results arising from quantum error correction~\cite{aharonov1996limitations,shor1996fault,aharonov1997fault,kitaev1997quantum,knill1998resilient}. 
We use $\widetilde{\text{QNC}}_1$ to denote the class of quantum circuits with depth $d = \mathcal{O}(\log n/\text{poly} \log \log n )$~\cite{aharonov1996limitations}, which is just below logarithmic.
We note that simpler classical simulations may be obtained for circuits with specific depths~\cite{rajakumar2024polynomial} or geometric locality, as we show, for example, in Theorem~\ref{thm sampling b}.} \label{result_table}
\end{table} 

The central framework of our approach is the decomposition of the Heisenberg time-evolved observable, $O$,  in the Pauli basis. 
Crucially, under uniform noise, a Pauli operator $P$ with weight $w[P]$  is damped by an amount $e^{-\gamma w[P]}$,  where the weight $w$ is defined as the number of non-identity elements in the operator. 
High-weight Pauli operators are thus almost entirely damped by noise, while low-weight operators are less affected. 
Since there are only polynomially many, $\mathcal{O}((3n)^w/w!)$, low-weight operators, this suggests that one might be able to efficiently approximate a time-evolved observable by only keeping track of such operators.
This intuition has motivated a range of seminal algorithms for classically simulating noisy quantum circuits by truncating Pauli operators of high weight~\cite{bremner2017achieving,gao2018efficient,barak2020spoofing,gao2021limitations,aharonov2023polynomial,yung2017can,fontana2023classical,shao2023simulating,tanggara2024classically,kuprov2007polynomially,karabanov2011accuracy,white2018quantum,ye2020emergent,rakovszky2022dissipation,von2022operator,white2023effective,yoo2023open,klein2022time,artiaco2024efficient,ermakov2024unified}.
However, whether it is possible to turn this intuition into a rigorous classical algorithm for general circuits has remained an essential open question.

There are two challenges. 
First, although each individual high-weight  operator is strongly damped by noise, there are, in principle,  exponentially many such operators.
To address this, existing classical algorithms~\cite{gao2018efficient,barak2020spoofing,gao2021limitations,bouland2022noise,aharonov2023polynomial,yung2017can,fontana2023classical,shao2023simulating} utilize a strongly simplifying property of \emph{random circuits}: In a random circuit, different Pauli paths \emph{do not coherently interfere} on average~\cite{fn3}.
This enables simple bounds on the error associated to truncated paths.
However, this property does not extend to general circuits.
Second, for circuits with gate-based noise, the damping of each Pauli operator is  given by $w_{U_t}[P]$, the number of non-identity elements of $P$ that are acted upon by gates in  $U_t$\footnote{\vspace{-10mm} Specifically, 
$w_{U_t}[P]$ counts the number of qubits that are both in the support of $P$ and acted on by non-identity gates in  $U_t$.}.
However, a high-weight operator might be involved only in a small number of gates in any given circuit layer.
Thus, in combination with the first challenge, any rigorous approach for bounding the truncation error \emph{must} keep track of the \emph{history} of operators across many circuit layers.

In what follows, we will begin by stating our main results (Theorems~\ref{thm 1} and~\ref{thm 2}). 
In particular, we provide one classical algorithm for circuits with uniform noise, which runs in polynomial-time for a large class of observables, and a second classical algorithm for circuits with gate-based noise, which runs in quasi-polynomial time.
We will then describe our algorithms, with a focus on delineating the key techniques and insights that allow us to overcome the two aforementioned challenges. 

\vspace{2mm}
\begin{theorem} \label{thm 1}
	{\emph{(Polynomial-time algorithm for quantum circuits with uniform noise)}}
	{Consider any noisy quantum circuit $\mathcal{C}$, any normalized observable $O$, and a low-average ensemble of states $\mathcal{E} = \{\rho\}$.
	Assume the Pauli coefficients of $O$ and $\rho$ can be efficiently computed.
	For circuits with uniform noise, there is a classical algorithm that computes the expectation values, $\tr( \mathcal{C} \{ \rho \} O)$,  in time 
	\begin{align}
            \emph{poly}(n) \cdot (1/\varepsilon)^{\mathcal{O}\left(\frac{\log(1/\gamma)}{\gamma^2}\right)}, & \hspace{1cm} \text{if $O$ is a sum of polynomially many Paulis} \label{eq: thm1 uniform} \\
        n^{\mathcal{O}\left(\frac{\log(1/\gamma)}{\gamma} \right) + \mathcal{O}\left(\frac{\log(1/\varepsilon)}{\gamma d}\right)} \cdot (1/\varepsilon)^{\mathcal{O}\left(\frac{\log(1/\gamma)}{\gamma^2}\right)}, & \hspace{1cm} \text{for any $O$}, \label{eq: thm1 uniform2}
	\end{align}
 \noindent with root-mean-square error $\varepsilon$ over the ensemble $\mathcal{E}$.
        }
\end{theorem}

\begin{theorem} \label{thm 2}
        {\emph{(Classical algorithm for quantum circuits with gate-based noise)}}
        {For circuits with gate-based noise, there is a classical algorithm  that runs in quasi-polynomial time,
        \begin{equation} \label{eq: thm1 gate}
	 	\mathcal{O} \left( d \cdot n^{\frac{1}{\gamma} \log(\sqrt{d+1}/\varepsilon)-1} \right).
	\end{equation}
  \noindent for any normalized $O$, with root-mean-square error $\varepsilon$ over the ensemble $\mathcal{E}$.}
 \end{theorem}

Before proceeding to our algorithms, a few brief comments are in order.
First, we say that an observable $O$ is normalized if its normalized Frobenius norm is equal to one, $\lVert O \rVert_F^2 \equiv \frac{1}{2^n} \text{tr}(  O^\dagger  O ) = 1$. For example, this is true for any Pauli operator. If an observable is not normalized, the errors $\varepsilon$ can be simply rescaled by the observable's norm.

Second, while our results apply to any noisy quantum circuit, they still involve a small amount of randomness via the ensemble of input states.
In particular, our metric of success is the root-mean-square error $\varepsilon$ over a \emph{low-average} ensemble of quantum states, which we define as any ensemble, $\mathcal{E} = \{ \rho \}$, whose mixture is close to the maximally mixed state, $\lVert \frac{1}{| \mathcal{E} |} \sum_\rho \rho \rVert_\infty \leq c/2^n$ for any $c = \mathcal{O}(1)$ (see Appendix~\ref{app: lemma 1}).
As an example, any complete basis of pure states, such as the computational basis, forms a low-average ensemble with $c=1$.
%
%
Intuitively, our attention to average-case performance excludes quantum error correction because syndrome qubits will be initialized randomly on average.
Experiments with random input states are of significant interest in several settings, including in quantum learning~\cite{caro2023out,huang2022learning} and quantum many-body dynamics~\cite{mi2022time,dupont2020universal,rosenberg2024dynamics}.

Third, if preferred, our bounds on the root-mean-square error $\varepsilon$ can  be easily translated to probability bounds.
That is: For any $\delta, \tilde{\varepsilon}$, our algorithms succeed in computing the expectation value $\tr(\mathcal{C}[\rho] O)$ to error $\tilde{\varepsilon}$ with high probability $1-\delta$ over input states $\rho$ drawn from $\mathcal{E}$.
This follows immediately from Theorems~\ref{thm 1} and~\ref{thm 2} by an application of Markov's inequality, setting $\varepsilon = \tilde{\varepsilon} \sqrt{\delta}$.
The runtime of the algorithms remains polynomial and quasi-polynomial, respectively, for any inverse-polynomial $\tilde{\varepsilon}$ and $\delta$.
Hence, we say that our algorithms succeed on ``most'' input states.

\begin{figure}
\noindent \begin{minipage}{0.492\textwidth}
\begin{algorithm}[H] 
    \centering
    \caption{Classical algorithm for uniform noise}\label{algorithm uniform}
    \vspace{1mm}
    \begin{flushleft}
    \textbf{Input:} State $\rho$,  circuit $\mathcal{C}$, observable $O$; threshold $\ell$ \\
    \textbf{Output:} Approximation of $\tr( \mathcal{C} \{ \rho \} O )$
    \end{flushleft}
    \vspace{-2.5mm}
    \begin{algorithmic}[1]
        \STATE{Enumerate Pauli paths $\vec{P}$ with $\sum_t w[P_t] \leq \ell$} 
        \STATE{Compute amplitudes $c_{\vec{P}} = \prod_t a^{(t)}_{P_{t-1}P_t}$} 
        \STATE{Apply noise $c_{\vec{P}} \gets e^{-\gamma \sum_t w[P_t]} \cdot c_{\vec{P}}$}
        \STATE{\textbf{return} $\sum_{\vec{P}} c_{\vec{P}} \tr( \rho P_d )$}
        \vspace{0.5mm}
    \end{algorithmic}
\end{algorithm}
\end{minipage}
\hfill
\begin{minipage}{0.492\textwidth}
\begin{algorithm}[H]
    \centering
    \caption{Classical algorithm for gate-based noise}\label{algorithm gate}
    \vspace{1mm}
    \begin{flushleft}
    \textbf{Input:} State $\rho$,  circuit $\mathcal{C}$, observable $O$; threshold $\ell$ \\
    \textbf{Output:} Approximation of $\tr( \mathcal{C} \{ \rho \} O )$
    \end{flushleft}
    \vspace{-2.5mm}
    \begin{algorithmic}[1]
        \STATE{Initialize $c_P = e^{-\gamma w[P]} \tr( O P )/2^n$ for $w[P] \leq \ell$} 
        \FOR{$t = 1$ to $d$} 
        	\STATE{Update via $c_Q \gets e^{-\gamma w_{U_t}[Q]} \sum_P a_{PQ}^{(t)} \cdot c_P $}
        \ENDFOR
        \vspace{-1.39mm}
        \STATE{\textbf{return} $\sum_P c_P \tr( \rho P )$} 
        \vspace{0.5mm}
    \end{algorithmic}
\end{algorithm}
\end{minipage}
\end{figure}

{\renewcommand\addcontentsline[3]{} \subsection{Classical algorithm for quantum circuits with uniform noise}}

Our algorithm for uniform noise (Algorithm~\ref{algorithm uniform}) estimates the expectation value, $\tr( \mathcal{C} \{ \rho \} O) = \text{tr}( \rho \, \mathcal{C}^\dagger  \{ O \} )$, by computing an approximation $\tilde{O}$ of the Heisenberg time-evolved observable $\mathcal{C}^\dagger \{ O \}$.
This approximation is given by the sum of all low-weight \emph{Pauli paths} that contribute to $\mathcal{C}^\dagger \{ O \}$.
Here, a Pauli path $\vec{P} = (P_0,\ldots,P_d)$ is a sequence of Pauli operators corresponding to each layer of the circuit [Fig.~\ref{fig: noisy circuit}(b)] and the exact time-evolved operator can immediately be expressed as a sum over Pauli paths, 
\begin{equation} \label{eq: pauli path decomposition}
    \mathcal{C}^\dagger \{ O \} = \sum_{\vec{P}} e^{-\gamma w[\vec{P}]} \left( \prod_{t=1}^d a^{(t)}_{P_{t-1} P_t} \right) P_d,
\end{equation}
with coefficients given by the transition amplitudes, $a_{PQ}^{(t)} = \frac{1}{2^n} \text{tr}( Q U_{t}^\dagger P U_t )$, at each layer $t$. 
Each path is damped by a factor $e^{-\gamma w[\vec{P}]}$ due to noise, where $w[\vec{P}] = \sum_t w[P_t]$ is the sum of the individual weights at each circuit layer.
Our approximation is given by the sum over all Pauli paths with weight below a threshold $\ell$,
\begin{equation}
    \tilde{O} = \sum_{\vec{P} : w[\vec{P}] \leq \ell} e^{-\gamma w[\vec{P}]} \left( \prod_{t=1}^d a^{(t)}_{P_{t-1} P_t} \right) P_d. \label{eq:tO}
\end{equation}
The estimated expectation value is $\text{tr}( \rho  \tilde{O} )$.

We hasten to emphasize that 
our algorithm is in fact identical to those from random circuits~\cite{gao2018efficient,aharonov2023polynomial}.
To this end,  our ability to rigorously extend the algorithm to general noisy quantum circuits owes to 
substantial improvements in a number of key proof techniques. 
In particular, to bound the error in the expectation values, we first prove the following simple lemma. 

\vspace{2mm}
\begin{lemma} \label{lemma: average error}
Consider a low-average ensemble $\mathcal{E} = \{ \rho \}$  and two observables $O$ and $\tilde{O}$.
The root-mean-square difference between the expectation value of $O$ and $\tilde{O}$ is less than $\lVert O - \tilde{O} \rVert_{F}$.
\end{lemma}
\noindent Thus, if we are able to bound the Frobenius norm of the sum of truncated Pauli paths, then we can bound the root-mean-square error in our algorithm.

To bound the norm, we group the exponentially many Pauli paths according to their weight at each circuit layer.
This organization is naturally viewed via a tree structure [Fig.~\ref{fig: proof}(a)].
Each layer $t$ of the tree corresponds to a layer of the quantum circuit, ordered from $t=d,\ldots,1$.
At every layer, each vertex branches into up to $n$ additional vertices, which correspond to Pauli paths with weight $w_t=1,\ldots,n$ at that layer. 
In this manner, each Pauli path $\vec{P}$ is uniquely associated with a particular downward sequence of vertices, $(w_d),(w_d,w_{d-1}),\ldots,(w_d,\ldots,w_0)$, through the  tree.

The key benefits of the Pauli tree are that: (i) the paths associated with each vertex are damped by at least $e^{-\gamma \sum_{t'=t}^d w_{t'}}$ due to noise, and (ii) the tree organization naturally enables us to bound the norm of sums of Pauli paths, by using the orthogonality of Pauli operators at each circuit layer.
We now outline how these properties enable a tight bound on the error of Algorithm~\ref{algorithm uniform}.
Let $\delta O(w,t)$ denote the sum of all Pauli paths from layer $0$ to $t$ with weight above $w$.
We are interested in the Frobenius norm of $\delta O(\ell,d)  = \mathcal{C}^\dagger \{ O \} - \tilde{O}$, which bounds the root-mean-square error of Algorithm~\ref{algorithm uniform} via Lemma~\ref{lemma: average error}.
To analyze the Frobenius norm, we leverage properties (i) and (ii) above to decompose it as a sum of contributions from each weight $w_d$ at the final circuit layer, $\lVert \delta O(\ell,d) \rVert_F^2 = \sum_{w_d} \lVert \mathcal{P}_{w_d} \{ \delta O(\ell,d) \} \rVert_F^2$, where $\mathcal{P}_{w_d}$ projects onto Pauli operators of weight $w_d$.
We can associate each term in the sum with a vertex, $(w_d)$, in the top layer of the Pauli tree.
We can then bound the contribution of each vertex as $\lVert \mathcal{P}_{w_d} \{ \delta O(\ell,d) \} \rVert_F^2 \leq e^{-2\gamma w_d} \lVert \delta O(\ell-w_d,d-1) \rVert_F^2$, since any Pauli path with weight $w_d$ at the final layer, and total weight above $\ell$, must have total weight above $\ell-w_d$ on the preceding layers (see Appendix~\ref{app: uniform} for details).
Combined with the first equality, we see that the norm of high-weight Pauli paths at layer $d$ is bounded by a sum of analogous norms at layer $d-1$.

To complete our proof, we recursively apply this bound down each layer of the Pauli tree.
That is, we bound the contribution of each top layer vertex, $(w_d)$, as a sum of contributions of vertices at the following layer, $(w_d,w_{d-1})$.
Each such contribution arises from Pauli paths with weight above $\ell-w_d-w_{d-1}$ on layers up to $d-2$, and incurs an associated damping, $e^{-2\gamma (w_d+w_{d-1})}$.
We then bound these contributions in terms of vertices at the next layer, $(w_d,w_{d-1},w_{d-2})$, and so on.
We halt each sequence of recursions once the summed weight, $\sum_{t'=t}^d w_{t'}$, surpasses $\ell-t$; from here, one can directly bound the remaining contribution by $e^{-2\gamma (\ell+1)}$, since every Pauli path from layer $0$ to $t-1$ has weight at least $t$~\cite{fn10}. 
In effect, this recursive process groups the truncated Pauli paths according to their earliest sub-sequence of weights that sum to above $\ell-t$.
Each grouping can be viewed as a single collective truncation by the algorithm [Fig.~\ref{fig: proof}(a)].
The total error of the algorithm is determined by the number of such truncations, which a simple counting argument shows is ${\ell \choose d}$.
This leads to a total error ${\ell \choose d} e^{-2\gamma(\ell+1)}$. We refer to Appendix~\ref{app: uniform} for full details.

From this error bound, we determine the runtime of Algorithm~\ref{algorithm uniform} by setting $\ell \approx \gamma^{-1} d + \gamma^{-1} \log(1/\varepsilon)$ to achieve a desired error  $\varepsilon$.
For circuits with uniform noise, the maximum non-trivial depth is $d = \mathcal{O}(\gamma^{-1} \log(1/\varepsilon))$, since after this depth, expectation values become $\varepsilon$-close to zero and the circuit can be trivially simulated~\cite{fn7}.
The runtime of our classical algorithm is determined by the number of Pauli paths with weight below $\ell$.
This is upper bounded by $n^{\ell/d} \cdot 2^{\mathcal{O}(\ell)}$ for any $O$~\cite{aharonov2023polynomial}, and $m \cdot 2^{\mathcal{O}(\ell)}$ when $O$ is a sum of $m$ Pauli operators (see Appendix~\ref{app: uniform}).
Plugging in the above values of $\ell$ and $d$ yields Theorem~\ref{thm 1}.

\vspace{-4mm}
{\renewcommand\addcontentsline[3]{} \subsection{Classical algorithm for quantum circuits with gate-based noise} }

Although Algorithm~\ref{algorithm uniform} is quite efficient for quantum circuits with uniform noise, where the depth $d$ is at most logarithmic, this efficiency does not extend to circuits with gate-based noise, where the depth $d$ may be large.
In such circuits, the runtime of Algorithm~\ref{algorithm uniform} diverges exponentially in $d$ owing to the ${\ell \choose d}$ truncations.

\FigProof

To address this, our algorithm for gate-based noise (Algorithm~\ref{algorithm gate}) performs only a single truncation at each circuit layer, of all Pauli operators with weight greater than $\ell$. 
In particular, at each layer $t$ we store an approximation of the time-evolved operator within the subspace of low-weight Pauli operators,
\begin{equation}
	\tilde{O}^{(t)} = \sum_{P : w[P] \leq \ell} \tilde{c}_P^{(t)} P.
\end{equation} 
The coefficients are updated from layer to layer via standard Heisenberg time-evolution,
\begin{equation} \label{eq: update coefficient}
	\tilde{c}_Q^{(t)} = e^{-\gamma w_{U_t}[Q]} \sum_{P : w[P] \leq \ell} a_{PQ}^{(t)} \, \tilde{c}_P^{(t-1)}. 
\end{equation}
At the final circuit layer, we compute the expectation value  $\text{tr}( \rho \tilde{O}^{(d)} )$.

By performing only $d+1$ truncations in total, Algorithm~\ref{algorithm gate} avoids the exponential blow-up of Algorithm~\ref{algorithm uniform} at large depths.
In particular, if we denote the truncated operators at layer $t$ as $\delta O^{(t)} \equiv \mathcal{P}_{> \ell} \{ \mathcal{D}_t \{ U_t^\dagger \tilde{O}^{(t-1)} U_t \} \}$, where $\mathcal{P}_{> \ell} = \sum_{w=\ell+1}^n \mathcal{P}_w$, then we immediately have $\lVert \delta O \rVert_F^2 \leq (d+1) \sum_{t=0}^d \lVert \delta O^{(t)} \rVert_F^2$ by the Cauchy-Schwarz inequality.
The trade-off is a moderate increase in runtime  compared to  Algorithm~\ref{algorithm uniform} at the same value of $\ell$.
In particular, Algorithm~\ref{algorithm gate} performs $d$ updates on $\mathcal{O}(n^{\ell})$ low-weight Pauli coefficients, and thus has runtime  $\mathcal{O}(d  n^{\ell})$; see Appendix~\ref{app: gate} for full details.

The central difficulty in proving Theorem~\ref{thm 2} is to bound the truncation norms, $\lVert \delta O^{(t)} \rVert_F$, within the gate-based noise model.
This would be straightforward within the uniform noise model, since a single application of uniform depolarizing noise damps the norm of any operator with weight above $\ell$ by at least $e^{-\gamma(\ell+1)}$.
However, as aforementioned, it is not clear whether such a bound  applies to the gate-based noise model, where the effect of noise can accumulate gradually over many circuit layers.  

To show that the bound indeed extends to circuits with gate-based noise, our core idea is to track the \emph{flow} of operator norm from one weight to another under each circuit gate [Fig.~\ref{fig: proof}(b)].
Intuitively, each two-qubit gate can transfer operator norm from a weight $w$ to  adjacent weights $w\pm 1$.
Thus, for a component of the operator to reach a high weight, it must participate in a high number of circuit gates, and thereby accrue a large damping due to noise.
Despite its simplicity, formalizing this intuition can be remarkably subtle, due to processes wherein an operator component grows to high weight, then shrinks back to low weight, and then grows to high weight again. 
Our focus on the flow of the operator norm allows us to simultaneously capture damping during the growth in operator weight, while avoiding  over-counting due to shrinking and re-growth processes.

In more detail, to quantify the norm of a time-evolved operator $O^{(g)}$ at weight $w$, we leverage an object of recent interest in many-body physics: the \emph{operator weight distribution}~\cite{roberts2018operator,schuster2023operator},
\begin{equation}
	P^{(g)}(w) = \frac{1}{2^n} \tr \big( O^{(g)} \cdot \mathcal{P}_{w} \{ O^{(g)} \} \big).
\end{equation}
Here, we track the evolution through each circuit gate $g$ instead of each circuit layer $t$.
In keeping with our intuition, we show that the weight distribution changes in a ``local'' manner when any unitary two-qubit gate $g$ is applied,
\begin{equation}
     P^{(g)}(w) = P^{(g-1)}(w) + J^{(g)}(w) - J^{(g)}(w+1),
\end{equation}
where $J^{(g)}(w)$ measures the flow in operator norm from weight $w-1$ to $w$ under  $g$.
The crucial insight of our proof is to show that this flow becomes \emph{lossy} for quantum circuits with gate-based noise, in the sense that 
\begin{equation} \label{eq: P change nice bound1}
	P^{(g)}(w) \leq P^{(g-1)}(w) + \{ 1, e^{-2\gamma} \} \cdot J^{(g)}(w) - \{ 1, e^{-2\gamma} \}\cdot J^{(g)}(w+1), \\
\end{equation}
where the bound holds for any choices of the constants $1$ or $e^{-2\gamma}$ in each of the terms.
The strongest bound is obtained by taking $e^{-2\gamma}$ when the term is positive, and $1$ when negative.
In the former case, the bound shows that increases in operator norm due to flows inward to weight $w$ are damped by $e^{-2\gamma}$ due to noise; in the latter case, decreases in the norm due to outward flows are not damped [Fig.~\ref{fig: proof}(b)].

To bound the truncation error, we apply Eq.~(\ref{eq: P change nice bound1}) for every circuit gate $g$ and sum over weights $w>\ell$.
The currents between high weights cancel, and we find that the total norm above weight $\ell$ is less than the \emph{net current} from $\ell$ to $\ell+1$, $\sum_g J^{(g)}(\ell)$, damped by $e^{-2\gamma}$.
By bounding the norm in terms of the net current, we naturally avoid subtleties due to shrinking and re-growth processes.
Furthermore, the net current from weight $\ell$ to $\ell+1$ is necessarily bounded by the maximum norm at weight $\ell$.
This is, by a similar inequality, less than the net current from $\ell-1$ to $\ell$ damped $e^{-2\gamma}$.
Iterating $\ell$ times, we find that the total norm above weight $\ell$ is less than $e^{-2\gamma(\ell+1)}$ for any  circuit with gate-based noise (see Appendix~\ref{app: gate} for full details).

\vspace{2mm}
\begin{lemma} \label{lemma: operator norm noise}
{\emph{(Upper bound on weight distributions in noisy quantum circuits)}}
Consider the time-evolution $O^{(g)}$ of any normalized observable $O$ in any quantum circuit with gate-based noise.
The cumulative weight distribution is less than
\begin{equation}
	\sum_{w'=w+1}^{\infty} P^{(g)}(w') \leq e^{-2\gamma(w+1)} 
\end{equation}
for all $\gamma, w, t$.
This also holds if we replace $O^{(g)}$ with our classical approximation $\tilde{O}^{(g)}$.
\end{lemma}
\noindent Applying the lemma to Algorithm~\ref{algorithm gate}, we find that the total truncation error is less than $(d+1) e^{-2\gamma (\ell+1)}$. 
Setting $\ell \approx \gamma^{-1} \,\text{log}(\sqrt{d}/\varepsilon)$ to ensure a small root-mean-square error gives Theorem~\ref{thm 2}.
We remark that our proof of Lemma~\ref{lemma: operator norm noise}, as outlined above, introduces substantially new techniques compared to any previous Pauli truncation algorithm that we are aware of.

\FigErrorMitigation


{\renewcommand\addcontentsline[3]{} \section{Applications and Implications} \label{sec: applications}}

\noindent Let us now consider the implications of our algorithms.
We summarize a few of the most prominent ones below, and direct the interested reader to the Appendix for details and additional results.

\vspace{3mm}
\noindent \textbf{Lower bound on quantum error mitigation}---Error mitigation seeks to estimate the output of an ideal quantum circuit via experiments on noisy circuits~\cite{cai2023quantum,li2017efficient,temme2017error,bonet2018low,mcardle2019error,rubin2018application,czarnik2021error,huggins2021virtual,o2021error}.
While the sampling overhead for certain mitigation strategies (such as probabilistic error cancellation) is  explicitly exponential in the number of qubits~\cite{temme2017error,li2017efficient,cai2023quantum}, $\sim \! e^{\gamma n d}$, for other strategies (such as zero noise extrapolation), an exponential scaling is only known for certain worst-case circuits~\cite{quek2022exponentially,takagi2022fundamental,takagi2023universal}.
This has raised an intriguing open question: Can error mitigation scale efficiently for any quantum circuits of practical interest~\cite{kim2023scalable,kim2023evidence,scholten2024assessing}?

Unfortunately, our results imply a somewhat negative answer.
For a large class of mitigation strategies, any circuit that can be error mitigated in polynomial time on most input states, can also be classically simulated in polynomial time on most input states (Appendix~\ref{app: error mit}). 
\vspace{2mm}
\begin{corollary}[Lower bound on error mitigation]
\label{cor: error mitigation}
Given an ideal circuit $\mathcal{C}$, Pauli observable $O$, and a low-average ensemble $\mathcal{E} = \{ \rho \}$.
Suppose an error mitigation strategy proceeds by measuring $O$ on any set of noisy circuits applied to $\rho$.
For any $\varepsilon = 1/\poly{n}$, if error mitigation can estimate $\tr( \mathcal{C} \{ \rho \} O )$ to root-mean-square error $\varepsilon/2$ in polynomial time,  there is a classical algorithm to compute $\tr( \mathcal{C} \{ \rho \} O )$ to root-mean-square error $\varepsilon$ in polynomial time. 
\end{corollary}
\noindent The reason is simple.
If mitigation can recover an ideal expectation value from noisy circuits, then one can also compute the value classically, by substituting our algorithm for the circuit results.
At an intuitive level, the properties that make error mitigation more efficient (having a signal dominated by low-weight Paulis) also make the signal easier to simulate.
We emphasize that this discussion applies only for noise rates that are constant with respect to $n$; for smaller noise rates, $\gamma = \tilde{\mathcal{O}}( 1/n )$, a circuit can be both complex and efficiently mitigable.

\vspace{3mm}
\noindent \textbf{Noisy DQC1 $\subseteq$ BPP}---Our results extend to circuits with a fixed input state within one   scenario: when the state is highly mixed. 
This follows directly from Theorems~\ref{thm 1},~\ref{thm 2} because the ensemble composed of a single highly mixed input state is low-average (Appendix~\ref{app: cor 1}).
Experiments with highly mixed input states are commonplace in several quantum technologies, including nuclear magnetic resonance spectroscopy~\cite{jones2010quantum} and solid-state quantum simulators~\cite{doherty2013nitrogen,zu2021emergent}.
Formally, this scenario is captured by the DQC1 model of computation~\cite{knill1998power}.
We prove that any noisy DQC1 computation can be classically simulated in polynomial time.

\vspace{3mm}
\noindent \textbf{Sampling from anti-concentrated circuits}---Our algorithms can also be used to sample from noisy quantum circuits, provided that the output distribution of the circuit \emph{anti-concentrates}.
This utilizes a standard reduction from noisy circuit sampling to  expectation values in the presence of anti-concentration~\cite{bremner2017achieving}.
We achieve sampling to within small total variational distance in quasi-polynomial time (Appendix~\ref{app: sampling}).

\vspace{3mm}
\noindent \textbf{Random circuits with non-unital noise}---Recent works have asked~\cite{fefferman2023effect,mele2024noise}: Can \emph{random} quantum circuits with \emph{non-unital} noise, such as spontaneous emission, escape the classical algorithms known for circuits with unital noise?
This was answered in part by Ref.~\cite{mele2024noise}, which provides an algorithm for expectation values in random circuits with non-unital noise, that runs in quasi-polynomial time for geometrically local circuits, and exponential time (in $\varepsilon^{-1}$) for general circuit architectures.
Leveraging our proof method for Theorem~\ref{thm 1}, we show that Algorithm~1 computes expectation values in \emph{polynomial} time for \emph{any} random circuit with non-unital noise (Appendix~\ref{app: non-unital}).
We note that a direct extension of Theorem~\ref{thm 1} to non-random circuits with non-unital noise is not possible, since for general quantum circuits with non-unital noise one can perform fault-tolerant computation~\cite{ben2013quantum}.

\vspace{3mm}
\noindent \textbf{Quantum advantage beyond sampling}---While the focus of our work has been on noisy quantum circuits, our results also have broader implications in the search for near-term experiments that exhibit a quantum advantage.
Namely, by inverting our Theorem~\ref{algorithm gate}, we show that any quantum circuit that is exponentially hard to  simulate classically \emph{must fail} in the presence of even a macroscopically small noise rate, $\gamma = \tilde{\Omega}(1/n)$  (Appendix~\ref{app: cor 3}). 
\begin{corollary} \label{cor: sensitivity}
{\emph{(Classically-hard quantum circuits must be highly sensitive to noise)}}
{Given any ideal quantum circuit $\mathcal{C}$, normalized observable $O$, and low-average ensemble $\mathcal{E} = \{ \rho \}$.
Suppose that the runtime of any classical algorithm to compute the expectation values, $\tr( \mathcal{C}\{ \rho\} O )$, to root-mean-square error $\varepsilon(n)$ is lower bounded by $\chi(n)$.
Then any noisy implementation of $\mathcal{C}$ \emph{must only succeed} in computing $\tr( \mathcal{C}\{ \rho\} O )$ to root-mean-square error $\varepsilon(n)$ for noise rates
\begin{equation}
	\gamma \leq \mathcal{O} \left( \frac{\text{\emph{log}}^2(n)}{\log(\chi(n))} \right).
\end{equation}
In particular, if the classical runtime is exponential in $n$, the noisy implementation must only succeed for macroscopically small noise rates, $\gamma = \mathcal{O}(\text{\emph{log}}^2(n)/n)$.
}
\end{corollary}
\noindent We propose that this criteria can function as a simple and rigorous ``test'' for whether a given quantum circuit can feature an advantage over classical computation.
Thus far, demonstrations of quantum advantage have been restricted to sampling experiments~\cite{arute2019quantum,wu2021strong}, which are by their nature almost maximally sensitive to noise~\cite{boixo2018characterizing,morvan2023phase}.
When moving beyond sampling, it can be difficult to design  experiments whose expectation values are macroscopically sensitive to noise in a non-trivial manner.
This is highlighted by recent experiments on IBM's quantum processor~\cite{kim2023evidence}, which, despite involving many qubits and high gate fidelities, were rapidly simulated classically by follow-up works~\cite{tindall2023efficient,beguvsic2023fast,kechedzhi2023effective,anand2023classical,rudolph2023classical,beguvsic2023fast,patra2024efficient}.
In light of our work, this can be understood as a consequence of the  experiments' insensitivity to noise~\cite{fn4}.

\noindent

{\renewcommand\addcontentsline[3]{} \section{Outlook} \label{sec: discussions}}

Our results provide the most extensive evidence yet, that  noisy quantum circuits without error correction can be efficiently classically simulated.
In the long-term, this emphasizes the necessity of quantum error correction for achieving a scalable quantum advantage over classical computation.
For experiments without error correction, our results demonstrate the necessity of improving the noise rate proportional to the number of qubits.
Moreover, even at sufficiently low noise rates, our results provide a rigorous test for whether a given  circuit can yield a quantum advantage, by establishing, perhaps counterintuitively, that such circuits must be highly sensitive to noise.

Our work opens the door to a number of intriguing  directions. 
First, is it possible to devise analogous classical algorithms for continuous-time dynamics instead of quantum circuits?
Second, to what extent do our results carry over to more general errors, such as dephasing or thermal noise?
Finally, can insights from our proof techniques be used to improve numerical algorithms for classically simulating quantum systems?
In particular, we speculate that our approach may provide a fruitful starting point for simulating highly-connected circuit architectures, where tensor network methods typically struggle.

\vspace{1mm}
\emph{Acknowledgements}---We are grateful to Dorit Aharonov, Zhenyu Cai, Matthias C. Caro, Andreas Elben, Bill Fefferman, Soumik Ghosh, Greg Kahanamoku-Meyer, John Preskill, Dominik Wild, and Mike Zalatel for valuable discussions and insights.
T.S. acknowledges
support from the Walter Burke Institute for Theoretical Physics at Caltech.
C.Y. is supported by the Department of Energy
under Quantum Pathfinder Grant DE-SC0024324.
X.G. acknowledges support from NSF PFC grant No. PHYS 2317149 and start-up grants from CU Boulder.
N.Y.Y acknowledges support from the NSF via the QLCI program (grant number OMA-2016245) and the STAQ II Program.
The Institute for Quantum Information and Matter, with which TS is affiliated, is an NSF Physics Frontiers Center.

\bibliographystyle{apsrev4-1} 
\bibliography{refs}

\begin{thebibliography}{115}
\bibitem [{fn1()}]{fn10}%
  \BibitemOpen
  \href@noop {} {\bibinfo  {journal} {At this halting step, we also group
  together all weights above the threshold in order to achieve a tighter bound
  on the error. See Appendix~C for full details}\ }\BibitemShut {NoStop}%
\bibitem [{fn8()}]{fn8}%
  \BibitemOpen
  \href@noop {} {\bibinfo  {journal} {To suppress the effect of read-out noise,
  one can consider implementing classical error correction prior to measurement
  (e.g.~via a repetition code on each qubit). In this case, the output
  distribution does not anti-concentrate and so our theorem does not apply}\
  }\BibitemShut {NoStop}%
\end{thebibliography}

\newpage
\widetext
\appendix

\tableofcontents

\section{Comparison to existing results} \label{app: comparison}

In this section, we provide a more detailed comparison between the runtime of  our classical algorithms and existing results.
Let us first briefly summarize this comparison for the gate-based noise model, and then turn to the case of uniform noise.

\begin{itemize}
    \item For quantum circuits with gate-based noise, the only existing classical algorithms are restricted to IQP circuits and random circuits (summarized in Table~I of the main text).
    Thus, for general circuits, Theorem~2 provides an exponential improvement over the naive classical simulation time, $2^{\mathcal{O}(n)}$.
\end{itemize}

This comparison becomes more involved for quantum circuits with uniform noise.
As for the gate-based noise model, the only existing classical algorithms for uniform noise which explicitly take advantage of the noise are specific to IQP circuits and random circuits.
However, since uniform noise limits the depth of any non-trivial quantum circuit~\cite{aharonov1996limitations}, we should compare the scaling of our theorems to existing algorithms for low-depth quantum circuits.
In particular, we show in Appendix~\ref{app: uniform} that expectation values in quantum circuits with uniform noise become $\varepsilon$-close to their values in the maximally mixed state at logarithmic depth, $d \geq \gamma^{-1} \log(1/\varepsilon)$~\cite{fn7}.
With this in mind, we compare the runtime of our classical algorithms to existing algorithms for circuits of such depth on a case-by-case basis below.

\begin{outline}
\1 Let us begin by considering general (i.e.~all-to-all-connected) circuit architectures. In this setting, there are no existing classical algorithms for logarithmic depth circuits, and so Theorem~1 provides an exponential improvement over the naive classical simulation time, $2^{\mathcal{O}(n)}$.
This exponential improvement continues to hold even for more restricted classes of observables, as we discuss below.

\2 For the specific case of local observables, we can moderately improve the naive classical simulation time by restricting to qubits within the light-cone of the observable. For general circuits, the light-cone contains $2^d$ qubits, leading to a runtime of $2^{\mathcal{O}(2^d)} = 2^{(1/\varepsilon)^{\mathcal{O}(\gamma^{-1})}}$. This is exponential in the inverse precision $\varepsilon^{-1}$, whereas our Algorithm~1 runs in polynomial time, $(1/\varepsilon)^{\tilde{\mathcal{O}}(\gamma^{-2})}$, for local observables.

\2 For tensor-product observables in general circuits, Ref.~\cite{bravyi2023classical} provides an algorithm with time $n^{\mathcal{O}(2^{2d} \log(n/\varepsilon))}$, i.e.~$n^{(1/\varepsilon)^{\mathcal{O}(\gamma^{-1})}  \log(n)}$, to estimate either (i) output probabilities, i.e.~$O = \dyad{0^n}$ or (ii) the magnitude of Pauli expectation values, $| \! \tr( \mathcal{C} \{ \rho \} O)|$, where $O$ is Pauli. The runtime is exponential in $\varepsilon^{-1}$ when applied to noisy quantum circuits, due to the double exponential dependence of the original runtime on the depth $d$. In contrast, in case (i), Algorithm~1 computes output probabilities to an \emph{exponentially smaller} error, $\varepsilon \cdot 2^{-n/2}$, in quasi-polynomial time. In case (ii), Algorithm~1 computes Pauli expectation values (including their sign) to within the same error as Ref.~\cite{bravyi2023classical} in polynomial time.

\1 We now turn to more restricted circuit architectures, beginning with the simplest case of 1D circuits. For tensor-product observables in 1D circuits, our algorithm provides no improvement because a trivial classical simuation via matrix product states is already very efficient.

\2 For tensor-product observables in 1D circuits, one can perform a classical simulation using matrix product states. This has runtime $2^{\mathcal{O}(d)}$, which translates to $(1/\varepsilon)^{\mathcal{O}(\gamma^{-1})}$ for quantum circuits with uniform noise. This is polynomial in $\varepsilon^{-1}$ as is our algorithm.

\1 For 2D circuits, existing classical algorithms become less efficient, and our results generally provide a quasi-polynomial to polynomial improvement.

\2 For local observables in 2D circuits, the best existing classical algorithm is a direct simulation restricted to the light-cone of the observable.
The light-cone contains $\mathcal{O}(d^2)$ qubits, which leads to a runtime  $2^{\mathcal{O}(d^2)} = (1/\varepsilon)^{\mathcal{O}(\gamma^{-2} \log(1/\varepsilon))}$ that is quasi-polynomial in $\varepsilon^{-1}$.
In contrast, our Algorithm~1 runs in polynomial time, $(1/\varepsilon)^{\tilde{\mathcal{O}}(\gamma^{-2})}$.
We note that we can also utilize the same light-cone argument to improve the scaling of our Algorithm~2, by replacing $n$ with $(\gamma^{-1} \log(1/\varepsilon))^2$ in Theorem~2.
This leads to a runtime $(1/\varepsilon)^{\mathcal{O}(\gamma^{-1}\log\text{log}(\varepsilon^{-1}))}$, which is nearly polynomial and has a better dependence on the noise rate $\gamma$.

\2 For tensor-product observables in 2D circuit architectures, Ref.~\cite{bravyi2021classical} provides an algorithm to estimate expectation values in time $\mathcal{O}(n d^2 2^{d^2} / \varepsilon^2)$ for any tensor-product observable, $O = \bigotimes_{j=1}^n O_j$ with $\lVert O_j \rVert_\infty \leq 1$. This translates to a runtime $n \cdot (1/\varepsilon)^{\mathcal{O}(\gamma^{-2} \log(1/\varepsilon))}$ for quantum circuits with uniform noise, which is quasi-polynomial in $\varepsilon^{-1}$. For Pauli observables, our algorithm improves the runtime to polynomial in $\varepsilon^{-1}$. 
For general tensor-product observables, we note that our algorithm computes expectation values to within precision $\varepsilon \cdot \lVert O \rVert_F$, where $\lVert O \rVert_F = \prod_{j=1}^n \lVert O_j \rVert_F$ may be substantially smaller than 1. 
Thus, replacing $\varepsilon \rightarrow \varepsilon/\lVert O \rVert_F$ in Theorem~1 for the sake of comparison to Ref.~\cite{bravyi2021classical}, our algorithm has runtime $n^{\tilde{\mathcal{O}}(\gamma^{-1}) + \mathcal{O}(\log(\lVert O \rVert_F/\varepsilon)/\gamma d)} (\lVert O \rVert_F/\varepsilon)^{\tilde{\mathcal{O}}(\gamma^{-2})}$ for general observables. 
This runtime may be better or worse than that of Ref.~\cite{bravyi2021classical} depending on the scalings of $n,d,\varepsilon,\gamma$.
For example, for large $n$ but moderate $\varepsilon$ and $d$, the algorithm in Ref.~\cite{bravyi2021classical} is more efficient than our algorithm.
On the other hand, for small $\varepsilon$ or large $d$ our algorithm becomes more efficient.
In particular, in the case $d = \Omega(\log(n))$,  our algorithm runs in polynomial time whereas the algorithm in Ref.~\cite{bravyi2021classical} runs in quasi-polynomial time.

\1 Finally, for 3D circuit architectures, Theorem~1 provides a sub-exponential to polynomial improvement over existing algorithms for tensor-product observables, and a quasi-polynomial to polynomial improvement for local observables.

\2 For tensor-product observables in 3D circuits, Ref.~\cite{bravyi2021classical} provides an algorithm with runtime,  $\mathcal{O}(n d^3 2^{d^2 n^{1/3}} \!\!/ \varepsilon^2)$, i.e.~$n \cdot (1/\varepsilon)^{\mathcal{O}(\gamma^{-2}  n^{1/3}\log(1/\varepsilon))}$ for quantum circuits with uniform noise. The runtime is sub-exponential in $n$, and is improved to either polynomial or quasi-polynomial time by our algorithm.

\2 For local observables in 3D circuits, a direct light-cone simulation runs in quasi-polynomial time,  $2^{\mathcal{O}(d^3)} = (1/\varepsilon)^{\mathcal{O}(\gamma^{-3} \log(1/\varepsilon)^2)}$.
In contrast, our Algorithm~1 runs in polynomial time, $(1/\varepsilon)^{\tilde{\mathcal{O}}(\gamma^{-2})}$, and our Algorithm~2 runs in nearly polynomial time, $(1/\varepsilon)^{\tilde{\mathcal{O}}(\gamma^{-1} \log \text{log} ( \varepsilon^{-1} ))}$.

\end{outline}

\section{Low-average ensembles of states} \label{app: lemma 1}

We now formally define our allowed ensembles of input states, which we call \emph{low-average ensembles}.
\vspace{2mm}
\begin{definition}
{\emph{(Low-average ensembles of states)}}
{We say that an ensemble of quantum states $\mathcal{E} = \{ \rho \}$ is a \emph{low-average ensemble with purity $c$} if $\lVert \frac{1}{| \mathcal{E} |} \sum_\rho \rho \rVert_\infty \leq c/2^n$.
For ensembles of pure states and $c=1$, this coincides with the definition of a 1-design of states.
}
\end{definition}
\vspace{2mm}
\noindent As an example, the ensemble of computational basis states with $n-m$ random bits and $m$ fixed bits forms a low-average ensemble with purity $c = 2^m$.
Any complete basis of pure states forms a low-average ensemble with $c=1$.

We now state and prove the more general version of our Lemma~\ref{lemma: average error} from the main text.
\vspace{2mm}
\begin{lemma} \label{lemma: average error app}
{\emph{(Extended version of Lemma~\ref{lemma: average error} of the main text)}}
Consider a low-average ensemble $\mathcal{E} = \{ \rho \}$ with purity $c$  and two observables $O$ and $\tilde{O}$.
The root-mean-square difference between the expectation value of $O$ and $\tilde{O}$ is less than $\sqrt{c} \cdot \lVert O - \tilde{O} \rVert_{F}$. 
\end{lemma}
\vspace{2mm}

\emph{Proof of Lemma~\ref{lemma: average error app}}---For simplicity, we begin by assuming that each state is pure, and afterwards generalize to mixed states. 
For a pure state $\rho = \dyad{\psi}$, the squared difference obeys
\begin{equation} \label{eq: pure state inequality}
	\tr(  \dyad{\psi} \delta O )^2 = \bra{\psi}  \delta O^\dagger  \dyad{\psi}  \delta O \ket{\psi_i} \leq \bra{\psi} \delta O^\dagger  \delta O \ket{\psi},
\end{equation}
where we recognize the middle expression as the expectation value of $\dyad{\psi}$ in the state $\delta O \ket{\psi}$, which is upper bounded by the normalization of the state.
Inserting this inequality into the mean-square error and applying the low-average condition gives our desired bound,
\begin{equation}
	\frac{1}{| \mathcal{E} |} \sum_\psi \tr( \dyad{\psi} \delta O )^2 \leq \frac{1}{| \mathcal{E} |} \sum_\psi \bra{\psi} \delta O^\dagger  \delta O \ket{\psi}  = \tr (  \rho_{\mathcal{E}}  \cdot \delta O^\dagger  \delta O ) \leq \lVert \rho_{\mathcal{E}} \rVert_\infty \cdot \lVert \delta O^\dagger \delta O \rVert_1 = c \cdot \lVert \delta O \rVert_{F}^2,
\end{equation}
where we use $\rho_{\mathcal{E}} = (1/|\mathcal{E}|) \sum_\psi \dyad{\psi}$ to denote the mixture of the ensemble.

We now turn to mixed states.
Let us write each mixed state in its orthonormal eigenbasis, $\{ \ket{ \psi_\rho } \}$, as $\rho = \sum_{\psi_\rho} p_{\psi_\rho} \dyad{\psi_\rho}$.
For each mixed state, we can define a quantum channel that dephases states in the eigenbasis of $\rho$, $\mathcal{D}_{\rho} \{ \dyad{\psi_\rho}{\psi_\rho'} \} = \delta_{\psi_\rho,\psi_\rho'} \dyad{\psi_\rho}$.
This allows us to write each mixed state $\rho$ as its corresponding dephasing channel $\mathcal{D}_\rho$ applied to a pure state $\ket{\phi_\rho}$,
\begin{equation}
	\rho = \mathcal{D}_{\rho} \{ \dyad{\phi_{\rho}} \}, \,\,\,\,\,\,\,\,\, \ket{\phi_{\rho}} = \sum_{\psi_\rho} \sqrt{p_{\psi_\rho}} \ket{\psi_\rho}.
\end{equation}
Turning to the mean-square difference in expectation values, we first re-write each trace as
\begin{equation}
    \tr( \rho \cdot \delta O) = \tr( \mathcal{D}_{\rho} \{ \dyad{\phi_{\rho}} \}  \cdot \delta O ) = \bra{\phi_{\rho}}  \mathcal{D}_{\rho} \{ \delta O \} \ket{\phi_{\rho}},
\end{equation} since the dephasing channel is equal to its conjugate.
Applying the same inequality as in Eq.~(\ref{eq: pure state inequality}) gives
\begin{equation} 
	\tr(  \rho \delta O )^2  \leq \bra{\phi_{\rho}} \mathcal{D}_{\rho} \{ \delta O^\dagger \}  \cdot \mathcal{D}_{\rho} \{ \delta O \} \ket{\phi_{\rho}}.
\end{equation}
Since both operators within the expectation value are diagonal in the eigenbasis of $\rho$, we can compute the expectation value as
\begin{equation}
	\bra{\phi_{\rho}} \mathcal{D}_{\rho} \{ \delta O^\dagger \}  \cdot \mathcal{D}_{\rho} \{ \delta O \} \ket{\phi_{\rho}} = \sum_{\psi_\rho} \sqrt{p_{\psi_\rho}} \cdot ( \delta O^\dagger )_{\psi_\rho \psi_\rho} \cdot ( \delta O )_{\psi_\rho \psi_\rho} \cdot \sqrt{p_{\psi_\rho}}
\end{equation}
This is upper bounded by the analogous expression that includes the off-diagonal matrix elements of $\delta O$,
\begin{equation}
	\sum_{\psi_\rho} \sqrt{p_{\psi_\rho}} \cdot ( \delta O^\dagger )_{\psi_\rho \psi_\rho} \cdot ( \delta O )_{\psi_\rho \psi_\rho} \cdot \sqrt{p_{\psi_\rho}} \leq \sum_{\psi_\rho,\psi_\rho'} p_{\psi_\rho} \cdot ( \delta O^\dagger )_{\psi_\rho \psi_\rho'} \cdot ( \delta O )_{\psi_\rho' \psi_\rho}  = \tr( \rho \cdot \delta O^\dagger \delta O ),
\end{equation}
since each off-diagonal element contributes an amount $p_{\psi_\rho} (\delta O^\dagger)_{\psi_\rho \psi_\rho'} (\delta O)_{\psi_\rho' \psi_\rho} = p_{\psi_\rho} | (\delta O)_{\psi_\rho' \psi_\rho} |^2 \geq 0$.
Putting it all together, we have
\begin{equation}
	\frac{1}{| \mathcal{E} |} \sum_\rho \tr(  \rho \delta O )^2 \leq \frac{1}{| \mathcal{E} |} \sum_\rho \tr( \rho \cdot \delta O^\dagger \delta O ) = \tr( \rho_\mathcal{E} \cdot \delta O^\dagger \delta O ) \leq c \cdot \lVert \delta O \rVert_F^2,
\end{equation}
where we denote $\rho_\mathcal{E} = \frac{1}{|\mathcal{E}|} \sum_\rho \rho$ once again. \qed

\section{Classical algorithm for uniform noise (Theorem~1)} \label{app: uniform}

We now provide our complete proof of Theorem~\ref{thm 1} of the main text.
We begin by establishing a short lemma that upper bounds the circuit depth of any non-trivial quantum circuit with uniform noise.
\vspace{2mm}
\begin{lemma} \label{lemma: uniform depth}
    {\emph{(Upper bound on circuit depth for quantum circuits with uniform noise)}}
    Consider any observable $O$ and any quantum circuit $\mathcal{C}$ with uniform noise.
    The Frobenius norm of the non-identity component of $\mathcal{C}^\dagger \{ O \}$ is upper bounded by $e^{-\gamma (d+1)} \lVert O \rVert_F$, where $d$ is the circuit depth.
\end{lemma}
\vspace{2mm}
\noindent \emph{Proof}---The Heisenberg time-evolution of the non-identity component of $O$ has weight $\geq 1$ at all layers. Thus, its Frobenius norm decreases by at least a factor $e^{-\gamma}$ from each noise channel. Since there are $d+1$ noise channels, we have $\lVert \mathcal{C}^\dagger \{ O \} - \tr(O)/2^n \cdot \mathbbm{1} \rVert_F \leq e^{-\gamma (d+1)} \lVert O \rVert_F$. \qed

The lemma can be viewed as an adaption of Theorem~3 in Ref.~\cite{aharonov1996limitations} to the Frobenius norm of observables.
Using our Lemma~\ref{lemma: average error app}, it implies that one can trivially simulate expectation values (to within small root-mean-square error $\varepsilon \cdot \sqrt{c} \cdot  \lVert O \rVert_F$ over a low-average ensemble of input states with purity $c$), in any quantum circuit with depth $d \geq  \gamma^{-1}\log(1/\varepsilon)$ and uniform noise.
To do so, one simply replaces the expectation value with its value in the maximally mixed state, $\text{tr}(O)/2^n$.
Thus, in what follows, we need only to analyze the performance of our Algorithm~1 on circuits of depth $d < \gamma^{-1}\log(1/\varepsilon)$.

\textbf{Accuracy of algorithm:} We establish the accuracy of Algorithm~1 by showing that our approximation $\tilde{O}$ is close to $O$ in Frobenius norm,
\begin{equation}\label{eq:tO-O<eps}
    \lVert \tilde{\OO} -\mathcal{C}^\dagger\{\OO\} \rVert_F\le \varepsilon \norm{\OO}_F,
\end{equation}
From Lemma~\ref{lemma: average error app}, this bounds the root-mean-square error in expectation values as desired.
For simplicity, we assume that the operator $O$ has no identity component; if it does, then this can be easily be included by adding a constant offset to the expectation values.

To show this, it is convenient to write operators as vectors in a doubled Hilbert space, with inner product $\iipp{A}{B} = \tr( A^\dagger B)/2^n$ for operators $A$, $B$.
The vector norm is equal to the squared Frobenius norm, $\iipp{A}{A} = \lVert A \rVert_F^2$.
In this notation, each unitary layer $U_t$ is represented by a super-unitary $\cU_t$ acting as $\cU_t \kket{A} \equiv \kket{U^\dagger_t A U_t}$. 
Meanwhile, the uniform noise channel acts on the basis of Pauli operators as $\mathcal{D} \kket{P} = e^{-\gamma w[P]} \kket{P}$.
Finally, the projector $\bP_\ell$ onto Pauli operators of weight $w$ is given by $\mathcal{P}_w \kket{P} = \delta_{w[P],w} \kket{P}$.
With this notation, our approximation $\tilde{O}$ in Eq.~(6) of the main text can be written concisely as \begin{equation}\label{eq:tO1}
    \tilde{\OO} = \sum_{w_0+\cdots+w_d\le \ell} e^{-\gamma(w_0+\cdots+w_d)} \bP_{w_d}\cU_d\cdots \bP_{w_2}\cU_2 \bP_{w_1}\cU_1\bP_{w_0} \keto{\OO},
\end{equation}
where $w_t\ge 1$ is the weight at each layer.
Each sequence of projectors in the sum corresponds to a single root-to-leaf sequence in the Pauli tree  [Fig.~\ref{fig: proof}(a) of the main text].

As discussed in the main text, to bound the error in our approximation, we break up the sum over $w_0,\ldots,w_d$ layer-by-layer, and use the fact that the $\mathcal{P}_{w_t}$ at each layer $t$ project onto orthogonal subspaces for different $w_t$.
In full detail, we have
\begin{align}\label{eq:O-OF<l1l2}
&\norm{\tilde{\OO} -\mathcal{C}^\dagger\{\OO\}}_F^2 = \norm{\sum_{w_0+\cdots+w_d>\ell} e^{-\gamma(w_0+\cdots+w_d)} \bP_{w_d}\cU_d\cdots \bP_{w_1}\cU_1\bP_{w_0} \keto{\OO} }^2 \nonumber\\
&\quad \le \sum_{w_d= 1}^{\ell-d+1} e^{-2\gamma w_d} \norm{\bP_{w_d}^{(\ell-d+1)} \, \cU_d\sum_{w_0+\cdots+w_{d-1}>\ell-w_d} e^{-\gamma(w_0+\cdots+w_{d-1})} \bP_{w_{d-1}}\cU_{d-1} \cdots  \bP_{w_1}\cU_1\bP_{w_0} \keto{\OO} }^2 \nonumber\\
&\quad \le \sum_{w_d= 1}^{\ell-d+1} e^{-2\gamma w_d} \norm{\sum_{ w_0+\cdots+ w_{d-1}>\ell- w_d} e^{-\gamma( w_0+\cdots+ w_{d-1})} \bP_{ w_{d-1}}\cU_{d-1}\cdots  \bP_{ w_1}\cU_1\bP_{ w_0} \keto{\OO} }^2 \nonumber\\
&\quad \le \sum_{ w_d= 1}^{\ell-d+1} \sum_{ w_{d-1}= 1}^{\ell - d +2 - w_d} e^{-2\gamma (w_d+ w_{d-1})} \norm{\bP_{ w_{d-1}}^{(\ell- d+2-w_d)}\cU_{d-1}\sum_{ \substack{w_0+\cdots+ w_{d-2} \\ >\ell- w_d- w_{d-1}}} e^{-\gamma( w_0+\cdots+ w_{d-2})} \bP_{ w_{d-2}}\cU_{d-2} \cdots \bP_{ w_1}\cU_1\bP_{ w_0} \keto{\OO} }^2 \nonumber\\
&\quad \le \cdots \le \sum_{ w_d= 1}^{\ell-d+1} \sum_{ w_{d-1}= 1}^{\ell-d+2- w_d} \cdots \sum_{ w_1=1}^{\ell- w_d-\cdots- w_2} e^{-2\gamma( w_1+\cdots+ w_{d})} \norm{\sum_{ w_0>\ell- w_d-\cdots- w_1} e^{-\gamma w_0} \bP_{ w_0} \keto{\OO}}^2 \nonumber\\
&\quad \le \cdots \le \sum_{ w_d= 1}^{\ell-d+1} \sum_{ w_{d-1}= 1}^{\ell-d+2- w_d} \cdots \sum_{ w_1=1}^{\ell- w_d-\cdots- w_2} e^{-2\gamma( \ell+1)} \norm{ \bP_{\ell+1-w_d-\cdots-w_1}^{(\ell+1-w_d-\cdots-w_1)} \keto{\OO}}^2 \nonumber\\
&\quad \le  e^{-2\gamma (\ell+1)} \norm{\OO}_F^2 \cdot \left( \sum_{ w_0+\cdots+ w_{d}=\ell+1} 1 \right).
\end{align}
Here, we let $\lVert \kket{A} \rVert^2 = \iipp{A}{A}$ denote the vector norm, and we define the modified projectors
\begin{equation}
    \bP^{(\ell)}_{w} \equiv \left\{\begin{array}{cc}
        \bP_{w}, & w<\ell, \\
        \bP_{\ell}+\bP_{\ell+1}+\cdots +\bP_n, & w=\ell,
    \end{array}\right.
\end{equation}
where $\mathcal{P}_{\ell}^{(\ell)}$ combines the projections onto all operators with weight above or equal to $\ell$.
In the second line of Eq.~(\ref{eq:O-OF<l1l2}), we use the fact that the $\bP^{(\ell-d+1)}_{ w_d}$ project onto orthogonal subspaces to write the total squared Frobenius norm as a sum of squared  norms on each subspace.
The third line follows because $\lVert \bP_{ w_d}^{(\ell-d+1)} \cU_d\keto{\OO'} \rVert \le \norm{\cU_d\keto{\OO'} }=\norm{\keto{\OO'} }$ for any operator $O'$, since $\bP_{ w_d}^{(\ell-d+1)}$ is a projector and $\cU_d$ a unitary.
The upper limit $\ell-d+1$ is chosen to be just large enough such that the resulting sum of weights is always at least $\ell+1$, since there are $d$ additional weights $w_{d-1},\ldots,w_0$ and each weight is at least $1$.
The fourth, fifth and sixth lines follow similarly.
In the second to last line, we use the fact that $ w_0 + \ldots +  w_d \geq \ell+1$.

The remaining sum in Eq.~(\ref{eq:O-OF<l1l2}) counts the number of sequences $(w_0,\ldots, w_d)$ that sum to $\ell+1$.
Each sequence corresponds to an individual truncation in our algorithm, as discussed and depicted in Fig.~\ref{fig: proof}(a) of the main text.
In mathematics, the number of such sequences is well-known (referred to as ``the number of compositions of $\ell+1$ into $d+1$ parts''), and is equal to ${\ell \choose d}$.
Thus, we have
\begin{align}
&\lVert \tilde{\OO} -\mathcal{C}^\dagger\{\OO\} \rVert_F^2 \leq e^{-2\gamma (\ell+1)} \cdot {\ell \choose d} \cdot \norm{\OO}_F^2. \\
\end{align}
To determine the required value of $\ell$, we use a standard bound on the binomial coefficient, ${\ell \choose d} \leq (e \ell/d)^d$.
We then have
\begin{align} \label{eq: binomial bound}
&  e^{-2\gamma (\ell+1)} \cdot {\ell \choose d} \leq e^{-2\gamma (\ell+1)} \cdot \left(\frac{e \ell}{d} \right)^d \cdot \norm{\OO}_F^2 \le e^{-\gamma (\ell+1)} \cdot \norm{\OO}_F^2,
\end{align}
where the second inequality holds as long as $d \log(e \ell/d) \leq \gamma (\ell+1)$.
Thus, we can ensure that the desired error bound Eq.~(\ref{eq:tO-O<eps}) holds by choosing \begin{equation}\label{eq:l=uniform}
    \ell = c_\gamma d + 2 \gamma^{-1} \log(1/\varepsilon),
\end{equation}
where $c_\gamma$ is the solution of $c_\gamma  = \gamma^{-1} \log(e c_\gamma)$; at small $\gamma$, we have $c_\gamma = \gamma^{-1} \log(e / \gamma) + \mathcal{O}(\gamma^{-1} \log\log(1/\gamma))$.

\textbf{Runtime of algorithm:} The runtime of Algorithm~1 is determined by the number of Pauli paths in our approximation of $\tilde{O}$. 
From Lemma~8 in Ref.~\cite{aharonov2023polynomial}, the number of such paths is upper bounded by \begin{equation}\label{eq:number_paulip}
    \text{min} \left( n^{\ell/d} , m \right) \cdot 2^{\mathcal{O}(\ell)} = \text{min} \left( n^{\mathcal{O}(c_\gamma + \frac{2}{\gamma d} \log(1/\varepsilon))} , m \right) \cdot 2^{\mathcal{O}(c_\gamma d)} \cdot (1/\varepsilon) ^{\mathcal{O}(1/\gamma)}
\end{equation} 
if the observable $O$ is a sum of $m$ Pauli operators.
To derive the second term in the minimum, we observe that in the second part of the proof of Lemma~8 of Ref.~\cite{aharonov2023polynomial}, we can choose the layer $t$ to be the final circuit layer, which by definition contains at most $m$ Pauli operators.
Otherwise we can choose the circuit layer with the minimum weight, which gives the first term in the minimum, $n^{\ell/d}$, as in Ref.~\cite{aharonov2023polynomial}.
Inserting the value of $c_\gamma$ and invoking  Lemma~\ref{lemma: uniform depth} to restrict attention to $d < \gamma^{-1} \log(1/\varepsilon)$, we find a runtime 
\begin{equation}\label{eq:number_paulip_2}
    \text{min} \left( n^{\mathcal{O}\left(\frac{\log(1/\gamma)}{\gamma} \right) + \mathcal{O}\left(\frac{\log(1/\varepsilon)}{\gamma d}\right)} , m \right) \cdot (1/\varepsilon)^{\mathcal{O}\left(\frac{\log(1/\gamma)}{\gamma^2}\right)}
\end{equation} 
Taking the first term in the minimum leads to the first scaling in Theorem~\ref{thm 1} of the main text. 
Taking the second term with $m = \text{poly}(n)$ leads to the second scaling. \qed

\section{Classical algorithm for gate-based noise  (Theorem~2)} \label{app: gate}

We now turn to our Algorithm~2 for gate-based noise. We begin in Appendix~\ref{app: lemma 2} with a proof of Lemma~\ref{lemma: operator norm noise} of the main text, and analyze the remaining aspects of the algorithm in Appendix~\ref{app: thm gate}.

\subsection{Bounding noisy operator growth} \label{app: lemma 2}

We now prove Lemma~\ref{lemma: operator norm noise} of the main text, which upper bounds the cumulative operator weight distribution,
\begin{equation}
	Q^{(t)}(w) = \sum_{w'=w+1}^n P^{(t)}(w'),
\end{equation}
in any noisy quantum circuit.
We also include a slightly stronger bound on the cumulative weight distribution when considering the approximate time-evolved operator $\tilde{O}$.
\vspace{2mm}
\begin{lemma} \label{lemma: operator norm noise app}
{\emph{(Extended version of Lemma~\ref{lemma: operator norm noise} of the main text)}}
Consider the Heisenberg time-evolution $O^{(t)}$ of an observable $O$ in a quantum circuit with gate-based noise.
The cumulative weight distribution is less than
\begin{equation}
	Q^{(t)}(w) \leq e^{-2\gamma(w+1)} \cdot \lVert O \rVert_F^2
\end{equation}
for all $\gamma, w, t$.
We can also extend this to our classical approximation $\tilde{O}$ of $O$, which truncates Pauli operators of weight greater than $\ell$ between circuit layers.
In this case, the cumulative operator weight distribution $\tilde{Q}^{(t)}(\ell+1)$ obeys
\begin{equation}
	\sum_{t = 0}^d \tilde{Q}^{(t)}(\ell) \leq e^{-2\gamma(\ell+1)} \cdot \lVert O \rVert_F^2.
\end{equation}
\end{lemma}
\vspace{2mm}

\emph{Proof of Lemma~\ref{lemma: operator norm noise app}}---We begin by proving the first statement; the second  will follow shortly after.
We keep track of the Heisenberg time-evolution of $O$ gate-by-gate.
We denote the operator evolved through the first $g$ gates and noise channels of the circuit as $O^{(g)}$; if $g_t$ denotes the final gate within a circuit layer $t$, then $O^{(g_t)}$ corresponds to $O^{(t)}$ in our layer-by-layer notation.
We again use $O^{(0)}$ to denote the operator evolved through the read-out noise channel.

The initial observable $O$ will have some operator weight distribution, $P(w)$, with norm $\lVert O \rVert_F^2$.
After the application of read-out noise, the weight distribution is damped to $P^{(0)}(w) = e^{-2\gamma w} P(w)$. 
This leads to a trivial upper bound on the cumulative operator weight distribution at time zero,
\begin{equation} \label{eq: upper bound Q0}
	Q^{(0)}(w) = \sum_{w' = w+1}^{N} e^{-2\gamma w'} P(w') \leq e^{-2\gamma (w+1)} \cdot Q(w) \leq e^{-2\gamma (w+1)} \cdot  \lVert O \rVert_F^2.
\end{equation}
Our aim is to show that this bound  in fact holds for all times.

We  proceed gate-by-gate.
The evolution during each gate consists of two steps: first, we apply the unitary gate $U_g$, then we apply local depolarizing noise to the two qubits in the gate.
In the first step, application of $U_g$ can transfer operator norm between adjacent values of $\ell$.
We capture this by defining the \emph{current}, $J^{(g)}(w)$, via
\begin{equation} \label{eq: Q change Ug}
	Q'^{(g)}(w-1) =  Q^{(g-1)}(w-1) + J^{(g)}(w),
\end{equation}
where $Q'^{(g)}(w-1)$ denotes the operator norm at weight above $ w-1$ after applying $U_g$ but before applying the noise channel $\mathcal{D}_g$.
Similarly, we have 
\begin{equation}
	P'^{(g)}(w) =  P^{(g-1)}(w) + J^{(g)}(w) - J^{(g)}(w+1),
\end{equation}
where $P'^{(g)}(w)$ is also defined after $U_g$ but before $\mathcal{D}_g$.
The current $J^{(g)}(w)$ quantifies the net transfer in operator norm from weight $w-1$ to $w$ by $U_g$.
Writing it out explicitly, we have
\begin{equation}
	J^{(g)}(w) =  \iipp{ O_{w-1,w} + O_{w,w} }{ O_{w-1,w}  + O_{w,w} }  - \iipp{ O_{w,w} }{ O_{w,w} } - \iipp{ O_{w,w-1} }{ O_{w,w-1} },
\end{equation}
where we use the doubled bra-ket notation to denote inner products of operators, $\iipp{ A }{B} \equiv \frac{1}{2^n} \tr( A B)$, and abbreviate $O_{w,w'} \equiv \mathcal{P}_{w'} \{ U^\dagger_g \mathcal{P}_{w} \{ O^{(g-1)} \} U_g \}$ for the component of the operator that starts at weight $w$ and ends at weight $w'$.
The first term is the contribution from weights $w-1$ and $w$ before the gate to the norm at weight $w$ after the gate.
The second and third term arise when taking the difference with $Q^{(g-1)}(w-1)$. 
Although it does not immediately appear so, the flow is in fact anti-symmetric upon exchanging $w-1$ and $w$, as it must be in order to preserve the total operator norm.
This can be verified by expanding the first term and using $\iipp{O_{w,w}}{O_{w-1,w}} = - \iipp{O_{w-1,w-1}}{O_{w,w-1}}$.
This equality is guaranteed because $U^\dagger_g \mathcal{P}_{w} \{ O^{(g-1)} \} U_g$ and $U^\dagger_g \mathcal{P}_{w-1} \{ O^{(g-1)} \} U_g$ are orthogonal.

To incorporate the noise channel $\mathcal{D}_g$, we use the fact that any operator that increases in weight under $U_g$ (in particular, the operator~$O_{w-1,w}$) must have support on both qubits in $U_g$ after the gate is applied.
Thus, the norm of such an operator is damped by a factor of $e^{-2\gamma}$ by the noise channel.
In full detail, we have
\begin{equation} \label{eq: P change original bound}
\begin{split}
	P^{(g)}(w) & = \bbra{ O_{w-1,w} + O_{w,w} + O_{w+1,w} } \mathcal{D}_g \cdot  \mathcal{D}_g \kket{ O_{w-1,w} + O_{w,w} + O_{w+1,w} } \\
 	& =  \bbra{ O_{w-1,w} + O^{1}_{w,w}  } \mathcal{D}_g \cdot  \mathcal{D}_g \kket{ O_{w-1,w} + O^{1}_{w,w} }
	+  \bbra{ O_{w+1,w} + O^{2}_{w,w}  }\mathcal{D}_g \cdot  \mathcal{D}_g \kket{ O_{w+1,w} + O^{2}_{w,w} } 
	\\
	& \,\,\,\,\,\,\,\,\,\,\, + \bbra{  O^3_{w,w}  }\mathcal{D}_g \cdot  \mathcal{D}_g \kket{O^3_{w,w}} \\
	& \leq e^{-4\gamma} \iipp{ O_{w-1,w} + O^{1}_{w,w}  }{ O_{w-1,w} + O^{1}_{w,w} }
	+ e^{-2\gamma} \iipp{ O_{w+1,w} + O^{2}_{w,w}  }{ O_{w+1,w} + O^{2}_{w,w} } 
	+ \iipp{  O^3_{w,w}  }{O^3_{w,w}} \\
	& = P^{(g-1)}(w) + e^{-4\gamma} \iipp{ O_{w-1,w} + O^{1}_{w,w}  }{ O_{w-1,w} + O^{1}_{w,w} }  - \iipp{ O^{1}_{w,w} }{ O^{1}_{w,w} } - \iipp{ O_{w,w-1} }{ O_{w,w-1} } \\
	& \,\,\,\,\,\,\,\,\,\,\, +  e^{-2\gamma} \iipp{ O_{w+1,w} + O^{2}_{w,w}  }{ O_{w+1,w} + O^{2}_{w,w} }   - \iipp{ O^{2}_{w,w} }{ O^{2}_{w,w} } - \iipp{ O_{w,w+1} }{ O_{w,w+1} } \\
	& \leq P^{(g-1)}(w) + \{ 1, e^{-4\gamma} \} \cdot \left( \iipp{ O_{w-1,w} + O^{1}_{w,w}  }{ O_{w-1,w} + O^{1}_{w,w} }  - \iipp{ O^{1}_{w,w} }{ O^{1}_{w,w} } - \iipp{ O_{w,w-1} }{ O_{w,w-1} } \right) \\
	& \,\,\,\,\,\,\,\,\,\,\, +  \{ 1, e^{-2\gamma} \} \cdot \left( \iipp{ O_{w+1,w} + O^{2}_{w,w}  }{ O_{w+1,w} + O^{2}_{w,w} }   - \iipp{ O^{2}_{w,w} }{ O^{2}_{w,w} } - \iipp{ O_{w,w+1} }{ O_{w,w+1} } \right) \\
 	& = P^{(g-1)}(w) + \{ 1, e^{-4\gamma} \} \cdot \left( \iipp{ O_{w-1,w} + O_{w,w}  }{ O_{w-1,w} + O_{w,w} }  - \iipp{ O_{w,w} }{ O_{w,w} } - \iipp{ O_{w,w-1} }{ O_{w,w-1} } \right) \\
	& \,\,\,\,\,\,\,\,\,\,\, +  \{ 1, e^{-2\gamma} \} \cdot \left( \iipp{ O_{w+1,w} + O_{w,w}  }{ O_{w+1,w} + O_{w,w} }   - \iipp{ O_{w,w} }{ O_{w,w} } - \iipp{ O_{w,w+1} }{ O_{w,w+1} } \right). \\
\end{split}
\end{equation}
In the second line, we perform the orthogonal decomposition $O_{w,w} = O^1_{w,w} + O^2_{w,w} + O^3_{w,w}$, where $O^1_{w,w}$ is proportional to $ O_{w-1,w}$, and $O^2_{w,w}$ is proportional to $ O_{w+1,w}$, and $O^3_{w,w}$ is orthogonal to both.
This is an orthogonal decomposition because $O_{w-1,w}$ and $O_{w+1,w}$ are orthogonal, which follows since $U^\dagger_g \mathcal{P}_{w-1} \{ O^{(g-1)} \} U_g$ and $U^\dagger_g \mathcal{P}_{w+1} \{ O^{(g-1)} \} U_g$ are orthogonal.
The three operators remain  orthogonal after application of $\mathcal{D}_g$, since $O_{w-1,w}$ and $O_{w+1,w}$ are eigenstates of the noise channel $\mathcal{D}_g$ (with eigenvalues $e^{-2\gamma}$ and $e^{-\gamma}$, respectively).
In the second to last line, we use that 
\begin{equation}
\begin{split}
    P^{(g-1)}(w) & = \iipp{ O_{w,w-1} + O_{w,w} + O_{w,w+1}}{ O_{w,w-1} + O_{w,w} + O_{w,w+1}} \\
    & = \iipp{ O_{w,w-1} }{ O_{w,w-1} } + \iipp{ O_{w,w+1} }{ O_{w,w+1} } + \iipp{ O^1_{w,w} }{ O^1_{w,w} }+ \iipp{ O^2_{w,w} }{ O^2_{w,w} }+ \iipp{ O^3_{w,w} }{ O^3_{w,w} },
\end{split}
\end{equation}
which follows since the five operators in the second line are orthogonal to one another.
Finally, in  the last line, we use the notation $\{ 1, e^{-4\gamma} \}$ and $\{ 1, e^{-2\gamma} \}$ to denote that the bound holds for either choice of constant, where one can make the choice independently between the two terms.
Recognizing the first three terms in parentheses as $J^{(g)}(w)$ and the latter three terms as $- J^{(g)}(w+1)$, we have the simpler bound,
\begin{equation} \label{eq: P change nice bound}
	P^{(g)}(w) \leq P^{(g-1)}(w) + \{ 1, e^{-4\gamma} \} \cdot J^{(g)}(w) - \{ 1, e^{-2\gamma} \}\cdot J^{(g)}(w+1), \\
\end{equation}
where again the bound holds for any choice of the constants.
When $J^{(g)}(w)$ is positive, the stronger bound is obtained by taking the second constant in the first term, and when it is negative, the first constant.
The reverse holds for the $J^{(g)}(w+1)$ term.
To summarize, increases in $P^{(g)}(w)$ are discounted by a factor $e^{-4\gamma}$ or $e^{-2\gamma}$ due to gate-based noise, while decreases are not.

We would like to utilize our update rule, Eq.~(\ref{eq: P change nice bound}), to bound the cumulative norm above a threshold weight $w$.
To begin, we can iterate Eq.~(\ref{eq: P change nice bound}) over each gate $h=1,\ldots,g$ in the circuit to obtain
\begin{equation} \label{eq: P change cumulative bound}
	P^{(g)}(k) \leq P^{(0)}(k) +  \left( \sum_{h=1}^g \{ 1, e^{-4\gamma} \} \cdot J^{(h)}(k) \right) - \sum_{h=1}^g J^{(h)}(k+1),
\end{equation} 
where we choose the constant $1$ for each term in the second sum, and for convenience change the variable from $w$ to $k$.
To bound the cumulative weight distribution above weight $w$, we can sum over $k = w+1, \ldots, N$. 
Taking the constant $e^{-4\gamma}$ for $k = w$, and the constant $1$ for all other $k$, we find
\begin{equation} \label{eq: Q change cumulative bound}
	Q^{(g)}(w) \leq Q^{(0)}(w) + e^{-4\gamma} \left( \sum_{h=1}^g J^{(h)}(w+1) \right). \\
\end{equation}
We would like to bound the sum within the parentheses above.
Intuitively, the sum cannot be too large, since the net current  from $w$ to $w+1$ is sourced from the operator's support on weight $w$.
To make this precise, let us observe Eq.~(\ref{eq: P change cumulative bound}) with $k = w$.
The rightmost term is precisely the sum we would like to bound.
Inverting the inequality, taking the constant $e^{-4\gamma}$ for each term in the first sum, and using that $P^{(g)}(w-1) \geq 0$, we have
\begin{equation}
	\left( \sum_{h=1}^g J^{(h)}(w+1) \right) \leq P^{(0)}(w) + e^{-4\gamma} \left( \sum_{h=1}^g J^{(h)}(w) \right).
\end{equation}
This bounds the net current from $w$ to $w+1$ by the initial norm on $w$, plus the net current from $w-1$ to $w$ discounted by $e^{-4\gamma}$.
We can iterate over the weight to obtain
\begin{equation}
	\left( \sum_{h=1}^g J^{(h)}(w+1) \right) \leq \sum_{k=1}^{w} e^{-4\gamma (k-1)} \cdot P^{(0)}(w+1-k).
\end{equation}
Inserting into Eq.~(\ref{eq: Q change cumulative bound}), we find
\begin{equation}
\begin{split}
	Q^{(g)}(w)  & \leq Q^{(0)}(w) + \sum_{k=1}^{w} e^{-4\gamma k} \cdot P^{(0)}(w+1-k).  \\
\end{split}
\end{equation}
The $k^{\text{th}}$ term in the sum corresponds to operators that begin at weight $w+1-k$ and evolve to weight greater than $ w$.
We find that such operators' contribution to $Q^{(g)}(w)$ is discounted by at least $e^{-4\gamma k}$, since they have participated in at least $k$ gates.
Expressing $P^{(0)}(w+1-k)$ and $Q^{(0)}(w)$ in terms of their values before read-out noise, we have
\begin{equation}
	Q^{(g)}(w)  \leq e^{-2\gamma w} \cdot Q(w) + \sum_{k=1}^{w} e^{-2\gamma k} \cdot e^{-4\gamma (w+1-k)}  \cdot P(w+1-k) \leq e^{-2\gamma (w+1)} \cdot \lVert O \rVert_F^2,
\end{equation}
as desired.
The right side can be improved to $e^{-4\gamma (w+1/2)} \cdot \lVert O \rVert_F^2$ if the initial observable has weight one. 

We now turn to the second inequality in the lemma.
Our classical algorithm truncates high-weight components of $\tilde{O}^{(t)}$ after each layer of the circuit.
As aforementioned, we denote the final gate in layer $t$ as $g_t$, so that the truncation occurs in between gates $g_t$ and $g_t+1$.
To begin, note that the truncation in our classical algorithm cannot increase the operator weight distribution.
This means that the upper bounds Eq.~(\ref{eq: upper bound Q0}) and Eq.~(\ref{eq: P change nice bound}) continue to apply to noisy quantum circuit with truncation.
Moreover, for the specific weight $w = \ell$, our upper bound on the cumulative weight distribution can be strengthened to
\begin{equation}
	\tilde{Q}^{(t)}(\ell) \leq e^{-4\gamma} \left( \sum_{h=g_{t-1}+1} ^{g_t} \tilde{J}^{(h)}(\ell+1) \right),
\end{equation}
since the norm above $\ell$ is reset to zero by the truncation after layer $t-1$.
Here, $\tilde{J}^{(g)}(\ell)$ is defined analogous to $J^{(g)}(\ell)$, but with $\tilde{O}$ instead of $O$.
Summing over the layers $t$, we have
\begin{equation}
	\sum_{t=0}^d \tilde{Q}^{(t)}(\ell) \leq \tilde{Q}^{(0)}(\ell) + e^{-4\gamma} \left( \sum_{h=1} ^{g_d} \tilde{J}^{(h)}(\ell+1) \right),
\end{equation}
where the right side is now the same quantity we bounded for the first inequality.
Repeating the steps of that proof, we find
\begin{equation}
	\sum_{t=0}^d \tilde{Q}^{(t)}(\ell) \leq e^{-2\gamma (\ell
 +1)} \cdot \lVert O \rVert_F^2,
\end{equation}
as desired. \qed

\subsection{Analysis of algorithm} \label{app: thm gate}

We now apply our bound on operator weight distributions to prove Theorem~2.
As for Theorem~1, we break our proof into two parts, which address the accuracy and runtime of Algorithm~2, respectively.

\vspace{2mm}
\textbf{Accuracy of algorithm:} Our classical algorithm truncates operator components with weight greater than $\ell$ in between each layer of the circuit.
Each truncation can increase the distance between the time-evolved operator $O^{(t)}$ and our approximation $\tilde{O}^{(t)}$ by at most
\begin{equation}
	\lVert O^{(t)} - \tilde{O}^{(t)} \rVert_F \leq \lVert O^{(t-1)} - \tilde{O}^{(t-1)} \rVert_F + \lVert \mathcal{P}_{> \ell} \{  \mathcal{D}_t \{ U_t^\dagger \tilde{O}^{(t-1)} U_t \} \} \rVert_F,
\end{equation}
by the triangle inequality.
The latter term is precisely the square root of $\tilde{Q}^{(t)}(\ell)$, the cumulative weight distribution for our approximation $\tilde{O}^{(t)}$. 
Summing over time steps $t= 0,\ldots,d$, we have
\begin{equation}
\begin{split}
	\lVert O^{(d)} - \tilde{O}^{(d)} \rVert_F & \leq \sum_{t=0}^d \sqrt{ \tilde{Q}^{(t)}(\ell) } \\
	& \leq \sqrt{d+1} \left( \sum_{t=0}^d \tilde{Q}^{(t)}(\ell) \right)^{1/2}  \\
	& \leq \sqrt{d+1} e^{-\gamma(\ell+1)} \lVert O \rVert_F,
\end{split}
\end{equation}
where in the second line we apply the Cauchy-Schwarz inequality, and in the third line we apply Lemma~\ref{lemma: operator norm noise app}.

From Lemma~\ref{lemma: average error app}, the above bound on the Frobenius norm of $O^{(d)} - \tilde{O}^{(d)}$ leads to an upper bound on the root-mean-square error for any low-average ensemble,
\begin{equation} \label{eq: eps ell}
	\left( \frac{1}{| \mathcal{E} |} \sum_\rho \big| \textrm{tr}( \rho O^{(d)} ) -  \textrm{tr}( \rho  \tilde{O}^{(d)} ) \big|^2 \right)^{1/2} \leq \sqrt{c} \, \sqrt{d+1} \cdot e^{-\gamma(\ell + 1)} \cdot \lVert O \rVert_F.
\end{equation}
For a desired error $\varepsilon \sqrt{c} \lVert O \rVert_F$, it suffices to take
\begin{equation} \label{eq: ell epsilon}
	\ell \geq \frac{1}{\gamma} \log \left( \frac{\sqrt{d+1}}{\varepsilon} \right)  - 1.
\end{equation}

\textbf{Runtime of algorithm:} We now turn to the complexity of classically computing the expectation value $\tr \big( \rho_i \tilde{O}^{(d)} \big)$.
As described in the main text, we do so by simulating the time-evolution of $\tilde{O}^{(t)} = \sum_{ P : w[P] \leq \ell } c_P^{(t)} P$ layer-by-layer in the Pauli basis.
This requires keeping track of the real-valued coefficients $c_P^{(t)}$ for all Pauli operators with weight less than $\ell$.
There are 
\begin{equation} \label{eq: D_ell}
	D_\ell = \sum_{k=0}^\ell {n \choose k} \cdot 3^k \leq {n \choose \ell} \cdot 3^{\ell} \cdot \frac{n-\ell+1}{n-\frac{4}{3}\ell +1} \leq \frac{3}{2} \frac{(3n)^\ell}{\ell!}
\end{equation}
such coefficients.
The sum counts the number Pauli operators with weight $k$ by first counting the number of sets of $k$ qubits, ${n \choose k}$, on which the Pauli has support, and then multiplying by the number of Pauli operators on those $k$ qubits, $3^k$.
The second inequality upper bounds the sum when $\ell \leq 3n/4$, and the final inequality upper bounds the fraction by $3/2$, which applies for $\ell \leq n/2$.

At the beginning of our algorithm, we initialize the coefficients $c_P^{(0)}$ by computing the Pauli coefficients of $O$.
As stated in the theorem, we assume that each coefficient can be computed efficiently, i.e.~in $\mathcal{O}(1)$ time\footnote{If the individual coefficients can be computed in time $\tau$ instead of time $\mathcal{O}(1)$, then the time to read-in all coefficients will be $\mathcal{O}(\tau \cdot D_\ell)$ instead of $\mathcal{O}( D_\ell)$. Our final runtime  will be unchanged as long as $\tau = \mathcal{O}(d \cdot \ell!)$, and will be quasi-polynomial in $n$ as long as $\tau$ is quasi-polynomial as well.}, and thus the entire vector in $\mathcal{O}(D_\ell)$ time.
At the end of our algorithm, we compute the trace $\text{tr}( \rho_i \tilde{O}^{(d)} )$ by computing the Pauli coefficients $\tr( \rho P)$ of $\rho$ and taking their inner product with $c_P^{(d)}$.
Again, if each coefficient can be computed in $\mathcal{O}(1)$ time, then the expectation value can be computed in $\mathcal{O}(D_\ell)$ time.

We update the coefficients from layer to layer via Eq.~(8) of the main text,
where the transition amplitudes $a_{QP}^{(t)}$ are computed from the depth-1 unitary $U_t = \prod_g U_g$ as
\begin{equation} \label{eq: full transition amplitude}
	a_{PQ}^{(t)} \equiv \frac{1}{2^n} \tr( Q U_t P U_t^\dagger ) = \prod_{g} \frac{1}{4} \tr_{\text{supp(g)}}( Q_{\text{supp(g)}} \cdot \mathcal{D}_g \{ U_g ^\dagger P_{\text{supp(g)}} U_{g} \} ).
\end{equation}
The product runs over all gates $g$ in $U_t$, where $\text{supp(g)}$ denotes the pair of qubits in gate $g$. 
We see that the $n$-qubit transition amplitudes factorize into a product of two-qubit transition amplitudes.
Each two-qubit amplitude can be computed in $\mathcal{O}(1)$ time, and thus the $n$-qubit amplitude in $\mathcal{O}(n)$ time.

A naive algorithm to update the coefficients would compute each transition amplitude $a_{QP}^{(t)}$ individually and thus require $\mathcal{O}(n \cdot D_\ell^2)$ time.
We can moderately improve this scaling by leveraging the sparsity of the transition amplitudes.
Note that the support of $U_t P U_t^\dagger$ consists of at most $2\ell$ qubits: the qubits in $\text{supp}(P)$ and those that they couple to under $U_t$.
Thus, for each $P$, we only need to compute transitions to $Q$ whose support lies within this set.
There are at most $(3/2) 3^\ell {2\ell \choose \ell} \leq (3/2) 12^\ell$ such $Q$, by a similar computation to Eq.~(\ref{eq: D_ell}).
Moreover, each transition amplitude can be computed from a product of only $\ell$ two-qubit transition amplitudes, in contrast to the naive number $n/2$.
Together, these lead to a runtime of $\mathcal{O}(D_\ell  \cdot  12^\ell \cdot \ell )$  per circuit layer.

Inserting our bound on $D_\ell$ and summing over the $d$ circuit layers, our algorithm requires
\begin{equation}
	\text{time} = \mathcal{O}\bigg( d  \cdot \ell \cdot \frac{( 36 n )^{\ell}}{\ell!} \bigg) = \mathcal{O}\left( d  \cdot n^{\ell} \right),
\end{equation} 
where $d$ is the depth of the circuit.
On the right, we simplify the expression by using  $\ell \cdot 36^{\ell} / \ell! = \mathcal{O}(1)$.
(For moderate values of $\ell \lesssim 36$, this may incur a large constant pre-factor in practice, although it will still be much smaller than the additional factor of $n^\ell$ in the naive update algorithm).
The algorithm uses space
\begin{equation}
	\text{space} = \mathcal{O}\bigg( \frac{(3 n)^\ell}{\ell!}  \bigg) = \mathcal{O}\big ( n^\ell \big),
\end{equation}
proportional to the number of coefficients $c_P^{(t)}$.
These requirements are comparable to performing exact time-evolution in a system of $n_{\text{eff}} \sim \ell \log(n)$ qubits.

For a desired error $\varepsilon$, we take $\ell$ according to Eq.~(\ref{eq: ell epsilon}).
Plugging into the above, this gives
\begin{equation} \label{eq: time}
	\text{time} =  \mathcal{O} \left( d \cdot n^{\frac{1}{\gamma} \log \left(\sqrt{d+1}/\varepsilon \right)-1} \right),
\end{equation} 
and
\begin{equation}
	\text{space} = \mathcal{O} \left( n^{\frac{1}{\gamma} \log \left(\sqrt{d+1}/\varepsilon \right) -1} \right).
\end{equation}
This concludes our proof. \qed

\section{Classical algorithms for sampling from noisy quantum circuits} \label{app: sampling}

In select cases, our classical algorithms can also be leveraged to sample from noisy quantum circuits.
In sampling, we are interested in reproducing the statistics when the final state of the circuit is measured in the computational basis. 
From Born's rule, each measurement produces an $n$-bit string $z$ with probability $p_{\rho,\mathcal{C}}(z) = \bra{z}  \mathcal{C} \{ \rho \} \ket{z}$.
In general, sampling is a strictly harder task than computing expectation values, since the distribution $p_{\rho,\mathcal{C}}(z)$ contains information about exponentially many expectation values in the computational basis.

We provide three algorithms for sampling from noisy quantum circuits.
The first two are direct extensions of Theorems~\ref{thm 1} and~\ref{thm 2}, and apply to any circuit on most input states.
Similar to previous works~\cite{bremner2017achieving,aharonov2023polynomial}, we require one additional assumption: that the output distribution of the circuit \emph{anti-concentrates}, in the sense that $(1/|\mathcal{E}|) \sum_\rho \sum_z p_{\rho,\tilde{\mathcal{C}}}(z)^2 = \mathcal{O}(2^{-n})$ for the ensemble $\mathcal{E} = \{ \rho \}$.
Here, $\tilde{\mathcal{C}}$ denotes the circuit  $\mathcal{C}$ with the read-out noise channel omitted.
Anti-concentration guarantees that most expectation values in the computational basis are close to zero.
In the presence of read-out noise, this enables one to sample from $p_{\rho,\mathcal{C}}(z)$ by computing only a small number of low-weight expectation values~\cite{bremner2017achieving}, which we show can be done efficiently using our Algorithms~\ref{algorithm uniform} and~\ref{algorithm gate} (see Appendix~\ref{app: sampling} for details).
\vspace{2mm}
\begin{subtheorem}{theorem}
\begin{theorem} \label{thm sampling a}
	{\emph{(Sampling from noisy quantum circuits; informal)}}
	{Consider a noisy quantum circuit $\mathcal{C}$ and a low-average ensemble $\mathcal{E} = \{ \rho \}$ whose Pauli coefficients can be efficiently computed.
        Assume the output distribution anti-concentrates, $(1/|\mathcal{E}|) \sum_\rho \sum_z p_{\rho,\tilde{\mathcal{C}}}(z)^2 = \mathcal{O}(2^{-n})$.
        Then for either uniform or gate-based noise, there is a classical algorithm to sample from $p_{\rho,\mathcal{C}}(z)$ within root-mean-square total variational distance $\varepsilon$ in quasi-polynomial time.
	}
\end{theorem}

Our third sampling algorithm specializes to quantum circuits with low depth.
In such circuits, one can exactly compute low-weight expectation values by restricting to the light-cone of the observable of interest.
Using this property, we show that one can classically sample from any low-depth circuit with \emph{read-out} noise, as long as the output distribution of the circuit anti-concentrates.
This holds even when the circuit gates themselves are noiseless.
\vspace{2mm}
\begin{theorem} \label{thm sampling b}
    {\emph{(Sampling from low-depth quantum circuits with read-out noise; informal)}}
	{Consider a quantum circuit $\mathcal{C}$ with depolarizing read-out noise, and a state $\rho$ whose Pauli coefficients can be efficiently computed.
        Assume the output distribution anti-concentrates, $\sum_z p_{\rho,\tilde{\mathcal{C}}}(z)^2 = \mathcal{O}(2^{-n})$.
        If the circuit has depth $d = \mathcal{O}(\log n)$ and is geometrically local, or has depth $d = \mathcal{O}(\log \log n )$ more generally, then there is classical algorithm to sample from $p_{\rho,\mathcal{C}}(z)$ within total variational distance $\varepsilon$ in quasi-polynomial time.
	}
\end{theorem}
\end{subtheorem}
\noindent Quantum advantages for sampling from low-depth circuits are highly sought, since low-depth circuits are generally less susceptible to experimental noise.
Nonetheless, our result shows that in many cases, even a small amount of read-out noise can preclude any such advantage~\cite{fn8}.

In the following subsections, we first establish the reduction from noisy circuit sampling to expectation values in the presence of anti-concentration, and then prove the theorems above.

\subsection{Reduction from noisy circuit sampling with anti-concentration to expectation values}

To extend our classical algorithms for expectation values to sampling, we leverage the following lemma.
The lemma follows almost entirely from Ref.~\cite{bremner2017achieving}, by adapting their Theorem~4 to state ensembles and neglecting the components of the theorem that are specific to IQP circuits.

\vspace{2mm}
\begin{lemma} \label{lemma: exp val to sampling}
{Consider any ensemble of states $\mathcal{E} = \{ \rho \}$, sampled in the computational basis with read-out noise $\gamma$ per qubit.
Let
\begin{equation}
        \overline{\alpha} = \frac{1}{|\mathcal{E}|} \sum_{\rho} 2^n \sum_{\textbf{s}} p_\rho(\textbf{s})^2
    \end{equation}
denote the mean collision probability of the output distribution before read-out noise, multiplied by $2^n$.
Suppose that, for some $\ell_s$, one can compute expectation values in $\rho$ of Pauli operators with weight less than $\ell_s$ to within root-mean-square error $\varepsilon'$, in time $\chi$.
Then there exists a classical algorithm to simulate sampling to within root-mean-square total variational distance
\begin{equation}
\varepsilon = 6\sqrt{2} \sqrt{ \varepsilon'^2  \, ( n^{\ell_s} + 1 ) + \bar{\alpha}  \, e^{-2\gamma \ell_s} }.
\end{equation}
The classical algorithm takes time $\mathcal{O}(\chi \cdot n^{\ell_s+1})$ per sample.}
\end{lemma}
\vspace{2mm}

\noindent The lemma also holds for any fixed state $\rho$, by taking the ensemble $\mathcal{E} = \{ \rho \}$ to be composed of only a single state.

\vspace{2mm}
\emph{Proof of Lemma~\ref{lemma: exp val to sampling}}---Our goal is to sample from the probability distribution \begin{equation}\label{eq:prhos}
    p_{\mathcal{D}\{\rho\} }(\textbf{s}) = \bra{\textbf{s}} \mathcal{D} \{ \rho \} \ket{\textbf{s}}. 
\end{equation}
Let us decompose the projector $\dyad{\textbf{s}}$ as a sum of Pauli operators,
\begin{equation}
    \dyad{\textbf{s}} = \frac{1}{2^n} \sum_{\textbf{t} \in \{0,1\}^n} (-1)^{\textbf{s}\cdot \textbf{t}} Z_{\textbf{t}} ,
\end{equation}
where $\textbf{s} \cdot \textbf{t} \equiv \sum_i s_i t_i$, and $Z_{\textbf{t}} = \bigotimes_{i=0}^n (Z_i)^{t_i}$ denotes the Pauli operator with identity support where $t_i = 0$ and $Z$ support where $t_i = 1$.
Since $w[Z_{\textbf{t}}] = |\textbf{t}|$, we have
\begin{equation} \label{eq: decompose p Zt}
    p_{\mathcal{D}\{\rho\} }(\textbf{s}) 
    = \frac{1}{2^n} \sum_{\textbf{t}} (-1)^{\textbf{s} \cdot \textbf{t}} \tr( Z_{\textbf{t}} \mathcal{D} \{ \rho \})
    = \frac{1}{2^n}\sum_{\textbf{t}} e^{-\gamma |\textbf{t}|} (-1)^{\textbf{s} \cdot \textbf{t}}  \tr( Z_{\textbf{t}} \rho ).
\end{equation}
This can be viewed as a Fourier transform over $\mathbbm{Z}_2^n$ of the coefficients $e^{-\gamma | \textbf{t}|} \tr( Z_{\textbf{t}} \rho)$. 

Following Ref.~\cite{bremner2017achieving}, our first step is to form a classical approximation, $q_{\mathcal{D}\{\rho\} }(\textbf{s})$, of the probability distribution $p_{\mathcal{D}\{\rho\} }(\textbf{s})$.
We do so via two steps.
First, we truncate all terms in Eq.~(\ref{eq: decompose p Zt}) with high-weight, $|\textbf{t}| > \ell_s$.
Second, we replace each remaining expectation value, $\tr( Z_{\textbf{t}} \mathcal{D} \{ \rho \})$,  with its classical estimate.
We denote the classical estimate as $a_{\textbf{t}}$; we have $(1/|\mathcal{E}|) \sum_{\rho} |a_{\textbf{t}} - \tr( Z_{\textbf{t}} \mathcal{D} \{ \rho \}) |^2 \leq \varepsilon'^2$ by assumption.
Together, these two steps give a classical approximation
\begin{equation}\label{eq:qrhos=}
    q_{\mathcal{D}\{\rho\} }(\textbf{s}) \equiv \frac{1}{2^n} \sum_{\textbf{t}:|\textbf{t}| \le \ell_s} (-1)^{\textbf{s} \cdot \textbf{t}} a_\textbf{t}. 
\end{equation}
The approximation may take small negative values, and is thus not guaranteed to be a probability distribution.
However, one can show that it is close to $p_{\mathcal{D}\{\rho\} }(\textbf{s})$ in mean-square total variational distance, 
\begin{align} \label{eq: tvd bound}
    \frac{1}{|\mathcal{E}|} \sum_{\rho} \left( \sum_{\textbf{s}} \left| p_{\mathcal{D}\{\rho\}(\textbf{s}) }-q_{\mathcal{D}\{\rho\} }(\textbf{s}) \right| \right)^2 &\le \frac{2^n}{|\mathcal{E}|} \sum_{\rho} \sum_{\textbf{s}} \lr{p_{\mathcal{D}\{\rho\} }(\textbf{s})-q_{\mathcal{D}\{\rho\} }(\textbf{s})}^2 \nonumber\\
    & = \frac{1}{|\mathcal{E}|} \sum_{\rho} \left( \sum_{|\textbf{t}|\leq \ell_s}  \left[ \tr( Z_{\textbf{t}} \mathcal{D} \{ \rho \})^2 - a_{\textbf{t}} \right]^2 + \sum_{|\textbf{t}| > \ell_s} e^{-2\gamma | \textbf{t} |} \tr( Z_{\textbf{t}} \rho )^2 \right) \nonumber\\
    & \leq \sum_{|\textbf{t}|\leq \ell_s}  \varepsilon'^2 +
    e^{-2\gamma (\ell_s+1)} \cdot \frac{1}{|\mathcal{E}|} \sum_{\rho} \sum_{|\textbf{t}| > \ell_s}  \tr( Z_{\textbf{t}} \rho )^2  \nonumber\\
    & \leq \sum_{|\textbf{t}|\leq \ell_s}  \varepsilon'^2 +
    e^{-2\gamma (\ell_s+1)} \cdot \frac{1}{|\mathcal{E}|} \sum_{\rho} \sum_{\textbf{t}}  \tr( Z_{\textbf{t}} \rho )^2  \nonumber\\
    & \leq \sum_{|\textbf{t}|\leq \ell_s}  \varepsilon'^2 +
    e^{-2\gamma (\ell_s+1)} \cdot \frac{1}{|\mathcal{E}|} \sum_{\rho} 2^n \sum_{\textbf{s}}  p_{\rho}(\textbf{s})^2  \nonumber\\
    & = (n^{\ell_s}+1)  \varepsilon'^2 +
    e^{-2\gamma (\ell_s+1)}\cdot \overline{\alpha} \nonumber\\
\end{align}
In going from the first to second line, we use Eqs.~(\ref{eq: decompose p Zt}) and~(\ref{eq:qrhos=}) to expand the first and second copy of $p_{\mathcal{D}\{\rho\}}(\textbf{s})$, $q_{\mathcal{D}\{\rho\}}(\textbf{s})$ as a sum over variables $\textbf{t}$ and $\textbf{t}'$.
Due to the sign factor $(-1)^{\textbf{s} \cdot \textbf{t}} (-1)^{\textbf{s} \cdot \textbf{t}'}$, the sum over $\textbf{s}$ gives a delta function $2^n \delta_{\textbf{t},\textbf{t}'}$.
Using the delta function to eliminate the sum over $\textbf{t}'$ gives the second line.
In going from the fourth to fifth lines, we reverse these steps to convert the second sum over $\textbf{t}$ to a sum over $\textbf{s}$.
The final line uses the fact that there are at most $n^{\ell_s}+1$ bitstrings with $|\textbf{t}| \leq \ell_s$, as well as the definition of the mean collision probability $\overline{\alpha}$.

To translate this approximation into a sampling algorithm, we use Lemma~10 in Ref.~\cite{bremner2017achieving}. 
The lemma shows that one can construct a true probability distribution, $\tilde{q}_{\mathcal{D}\{\rho\} }(\textbf{s})$, from $q_{\mathcal{D}\{\rho\} }(\textbf{s})$, at the cost of only a slightly larger total variational distance,
\begin{equation}
    \sum_{\textbf{s}} \left| p_{\mathcal{D}\{\rho\} }(\textbf{s})-\tilde{q}_{\mathcal{D}\{\rho\} }(\textbf{s}) \right| \le 4\eta_\rho/(1-\eta_\rho),\quad \mathrm{where}\quad \eta_\rho \equiv \sum_{\textbf{s}} \left| p_{\mathcal{D}\{\rho\} }(\textbf{s})-q_{\mathcal{D}\{\rho\} }(\textbf{s}) \right|,
\end{equation}
for each $\rho$.
We can translate this to the mean-square total variational distance using
\begin{align}
    \frac{1}{|\mathcal{E}|} \sum_{\rho} \left( \sum_{\textbf{s}} \left| p_{\mathcal{D}\{\rho\} }(\textbf{s})-\tilde{q}_{\mathcal{D}\{\rho\} }(\textbf{s}) \right| \right)^2 &\le \frac{1}{|\mathcal{E}|} \sum_{\rho:\eta_\rho>1/3} 2^2 + \frac{1}{|\mathcal{E}|} \sum_{\rho: \eta_\rho\le 1/3} \lr{\frac{4\eta_\rho}{1-\eta_\rho}}^2 \nonumber\\
    &\le 2^2 \frac{\varepsilon^2}{(1/3)^2} + 6^2 \frac{1}{|\mathcal{E}|} \sum_{\rho: \eta_\rho\le 1/3} \eta_\rho^2 \le 72 \varepsilon^2,
\end{align}
where we denote the right hand side of the final line of Eq.~(\ref{eq: tvd bound}) as $\varepsilon^2$.
In the first line, we use that the total variational distance is upper bounded by $2$ for any two probability distributions, which allows us to replace $4\eta_\rho/(1-\eta_\rho)$ by $2$ when $\eta_\rho > 1/3$.
In the second line, we use Markov's inequality to bound the probability that $\eta_\rho > 1/3$.

To sample from $\tilde{q}_{\mathcal{D}\{\rho\}}(\textbf{s})$, we invoke Section 3.2 of Ref.~\cite{bremner2017achieving}.
The section provides a simple sampling algorithm, given the ability to compute marginals of the original approximation $q_{\mathcal{D}\{\rho\} }(\textbf{s})$. Any such marginal can be computed exactly by expanding $q_{\mathcal{D}\{\rho\} }(\textbf{s})$ as in Eq.~(\ref{eq:qrhos=}), \begin{equation}
    \sum_{\textbf{s}:\textbf{s}_{1\cdots k}=\textbf{y}} q_{\mathcal{D}\{\rho\} }(\textbf{s}) = \frac{1}{2^n} \sum_{|\textbf{t}| \le \ell_s} \sum_{\textbf{s}:\textbf{s}_{1\cdots k}=\textbf{y}} (-1)^{\textbf{s} \cdot \textbf{t}} a_P = \frac{1}{2^n} \sum_{|\textbf{t}| \le \ell_s} \delta_{\textbf{t}_{k+1 \cdots n},\mathbf{0}} \cdot (-1)^{\textbf{y} \cdot \textbf{t}_{1\cdots k}} \cdot a_P.
\end{equation}
Here, the left side is the marginal distribution on the first $k$ qubits, where $\textbf{y}\in\{0,1\}^k$. 
To compute the marginal, one can enumerate each low-weight $\textbf{t}$ and compute the sum on the right side exactly term-by-term.
This requires time $\chi (n^{\ell_s}+1)$.
Generating a single sample requires that we compute one marginal for each value of $k=1,\ldots,n$~\cite{bremner2017achieving}, and thus requires time $\chi (n^{\ell_s}+1) n$. \qed
\vspace{1.5mm}

\subsection{Proofs of sampling algorithms}

We now provide the detailed statements and proofs of our theorems on sampling from noisy quantum circuits.
As before, we let $\tilde{\mathcal{C}}$ denote the circuit  $\mathcal{C}$ with the read-out noise channel omitted.
For general circuits, we have
\vspace{2mm}
\begin{subtheorem}{theorem}
\begin{theorem} \label{thm sampling a app}
	{\emph{(Sampling from noisy quantum circuits; formal)}}
	{Consider a noisy quantum circuit $\mathcal{C}$ and a low-average ensemble $\mathcal{E} = \{ \rho \}$, where the Pauli coefficients of each $\rho$ can be efficiently computed.
        Let $\overline{\alpha}$ denote the mean collision probability of $\tilde{\mathcal{C}}\{ \rho \}$ multiplied by $2^n$.
        For circuits with uniform noise, there is a classical algorithm to sample from $\mathcal{C}\{ \rho \}$ to within root-mean-square total variational distance $\varepsilon$ that runs in time
        \begin{equation} \label{eq: thm2 uniform}
	 	n^{\mathcal{O}(\gamma^{-3} \log(1/\gamma) \log(\overline{\alpha}/\varepsilon))}.
	\end{equation}
        For circuits with gate-based noise, there is a classical algorithm that runs in time
        \begin{equation} \label{eq: thm2 gate}
	 	n^{\mathcal{O}( \gamma^{-2} \log(n) \log( \overline{\alpha} /\varepsilon)   + \gamma^{-1} \log( d ))}.
	\end{equation}
	}
\end{theorem}
\vspace{2mm}
\noindent In both cases, the runtime is quasi-polynomial so long as $\overline{\alpha} = \text{poly}(n)$.
For simplicity of presentation, in the main text we take $\overline{\alpha} = \mathcal{O}(1)$, which is referred to as \emph{anti-concentration}~\cite{bremner2017achieving,aharonov2023polynomial}.

\emph{Proof of Theorem~\ref{thm sampling a app}}---We take $\ell_s = \frac{1}{\gamma} \log( \sqrt{2 \bar{\alpha}} / \varepsilon )$ in Lemma~\ref{lemma: exp val to sampling} to ensure that the second term in the error is small.
This requires $2\varepsilon'^2 \leq \varepsilon^2/ (n^{\ell_s}+1)$ for the first term.
For circuits with uniform noise, we can compute the expectation value of any Pauli operator in time $\chi = (1/\varepsilon')^{\mathcal{O}\left(\frac{\log(1/\gamma)}{\gamma^2}\right)}$ (Theorem~1).
From Lemma~\ref{lemma: exp val to sampling}, this gives a total runtime
\begin{equation}
    \mathcal{O}(\chi\cdot n^{\ell_s+1}) = \left(  n^{\ell_s} / \varepsilon^2 \right)^{\mathcal{O}\left(\frac{\log(1/\gamma)}{\gamma^2}\right)} =  n^{\mathcal{O}(\gamma^{-3} \log(1/\gamma) \log(\overline{\alpha}/\varepsilon))},
\end{equation}
which establishes Eq.~(\ref{eq: thm2 uniform}).
Similarly, for circuits with gate-based noise, setting $\chi$ according to Eq.~(\ref{eq: time}) gives a total runtime 
\begin{equation}
	n^{\mathcal{O} \left( \frac{1}{\gamma} \log( d n^{\ell_s} / \varepsilon ) \right)} = n^{\mathcal{O}( \gamma^{-2} \log(n) \log( \bar{\alpha} /\varepsilon) + \gamma^{-1} \log( d ) )}.
\end{equation}
which establishes Eq.~(\ref{eq: thm2 gate}). \qed
\vspace{1.5mm}

For low-depth circuits, we have
\vspace{2mm}
\begin{theorem} \label{thm sampling b app}
    {\emph{(Sampling from low-depth quantum circuits with read-out noise; formal)}}
	{Consider any  circuit $\mathcal{C}$ that is either depth $d = \mathcal{O}(\log(n))$ and geometrically-local, or depth $d = \mathcal{O}(\log(\log(n))$, and contains read-out noise $\gamma$ per qubit.
        Let $\rho$ be any state whose Pauli coefficients can be efficiently computed, and $\alpha$ denote the collision probability of $\tilde{\mathcal{C}}\{ \rho \}$  multiplied by $2^n$. 
        There is classical algorithm to sample from $\mathcal{C} \{ \rho \}$  within total variational distance $\varepsilon$ in  time
        \begin{equation} \label{eq: thm2 low-depth}
	 	2^{\mathcal{O}( \gamma^{-1} \log( \overline{\alpha} / \varepsilon ) (\log n)^{c}  )},
	\end{equation}
 where $c=D$ for geometrically-local circuits in spatial dimension $D$, and $c = \mathcal{O}(1)$ for general circuits.
	}
\end{theorem}
\end{subtheorem}
\vspace{2mm}

\emph{Proof of Theorem~\ref{thm sampling b app}}---The theorem follows from a simple observation: that expectation values of low-weight Pauli operators can be computed exactly in any  circuit of sufficiently low depth.
More precisely, any Pauli operator with initial weight at most $\ell_s$ can have support on at most $\mathcal{O}(\ell_s \cdot d^D)$ or $\mathcal{O}(\ell_s \cdot 2^d)$ qubits after Heisenberg time-evolution. 
The first scaling applies to geometrically-local circuits, and the second to any circuit.
The expectation value of the Pauli operator can therefore be computed exactly in time $2^{\mathcal{O}(\ell_s \cdot d^D)}$ or $2^{\mathcal{O}(\ell_s \cdot 2^d)}$.
Therefore, setting $\varepsilon' = 0$ and $\ell_s = \gamma^{-1} \log(6\sqrt{2\overline{\alpha}}/\varepsilon)$ in Lemma~\ref{lemma: exp val to sampling}, one can classically sample from the circuit in time $2^{\mathcal{O}(\ell_s \cdot d^D) + \mathcal{O}(\ell_s \cdot \log(n))}$ or $2^{\mathcal{O}(\ell_s \cdot 2^d) + \mathcal{O}(\ell_s \cdot \log(n))}$.
Setting $d = \mathcal{O}(\log(n))$ for geometrically-local circuits, the first scaling becomes $2^{\mathcal{O}(\ell_s \cdot \log(n)^D)} = 2^{\mathcal{O}(\gamma^{-1} \log(\overline{\alpha}/\varepsilon) \cdot \log(n)^{D})}$.
Meanwhile, setting $d = \mathcal{O}(\log(\log(n)))$ for general circuits, the second scaling becomes $2^{\ell_s \cdot \log(n)^{\mathcal{O}(1)}}=2^{\gamma^{-1} \log(\overline{\alpha}/\varepsilon) \cdot \log(n)^{\mathcal{O}(1)}}$.
This concludes our proof. \qed

\section{Extension to circuits with noise only in non-Clifford gates} \label{app: non-clifford}

In this section, we show that both of our classical algorithms naturally extend to settings where noise occurs only in non-Clifford gates, and Clifford gates are implemented noiselessly.
Intuitively, this follows because our algorithms track operators in the Pauli basis.
Clifford gates  permute the Pauli operators, and thus do not increase the complexity of our classical approximation.
We formalize this in the following extension of our main result.
\vspace{2mm}
\begin{extension} \label{ext: non-clifford}
{\emph{(Classical algorithm for noisy quantum circuits with non-Clifford noise)}}
Consider the settings of Theorems~\ref{thm 1},~\ref{thm 2} of the main text but where (i) for the uniform noise model, noise occurs only in circuit layers that contain non-Clifford gates and during read-out, and (ii) for the gate-based noise model, noise occurs only in non-Clifford gates and during read-out.
Then all of the stated results continue to hold, with the scaling in Theorem~2 replaced by 
\begin{equation} \label{eq: non-clifford gate}
    \mathcal{O} \left( d \cdot m^2 \cdot n^{\frac{2}{\gamma} \log(\sqrt{d+1}/\varepsilon) } \right),
\end{equation}
where $d$ is the number of circuit layers containing non-Clifford gates, and $m$ is the number of Pauli operators with non-zero coefficients in $O$ and weight less than or equal to $\ell$.
\end{extension}
\vspace{2mm}
\noindent The result continues to hold even if the read-out is noiseless, provided that $O$ is a sum of polynomially many Pauli operators, $m = \text{poly}(n)$.

\emph{Proof of Extension~\ref{ext: non-clifford}}---We provide separate proofs for our algorithms for uniform and gate-based noise

\textbf{Uniform noise: }Let us begin with our algorithm for uniform noise. We use the same approximation, $\tilde{\OO}$, of the Heisenberg time-evolved operator as in Eq.~(\ref{eq:tO1}), but where $\cU_t$ now corresponds to the unitary of the $t$-th non-Clifford layer combined with all of the ensuing Clifford layers before the $(t+1)$-th non-Clifford layer. 
One can verify that Eq.~(\ref{eq:O-OF<l1l2}) still holds, because it only relies on the super-unitarity of $\cU_t$. 
Therefore, as before, the approximation $\tilde{\OO}$ is accurate with small Frobenius error [Eq.~(\ref{eq:tO-O<eps})] provided that we choose the truncation weight $\ell$ as in Eq.~(\ref{eq:l=uniform}).
Viewing Eq.~(\ref{eq:l=uniform}), the depth $d$ now corresponds to the number of non-Clifford layers, and, as before, we have $d < \gamma^{-1} \log(1/\varepsilon)$ before the system approaches the maximally mixed state.

It remains to show that our algorithm remains efficient---i.e.~that the scaling of the number of Pauli paths in Eq.~(\ref{eq:tO1}) is unchanged.
Following Ref.~\cite{aharonov2023polynomial}, we count the number of Pauli paths for each fixed sequence $(\ell_0,\cdots,\ell_d)$ individually, where $\ell_0+\ldots+\ell_d \leq \ell$, since there are at most $2^\ell$ such sequences.
One starts from the non-Clifford layer with the smallest weight $\ell_t$, at the beginning of which there are at most $(3n)^{\ell/d}$ possible Pauli operators (since one must have $\ell_t \leq \ell/d$). 
If $O$ is a sum of polynomially many Pauli operators, we start instead at the final circuit layer $t=0$, in which case this factor is replaced by $m$.
The ensuing non-Clifford layer can ``split'' each Pauli operator to at most $2^{\mathcal{O}(\ell_t)}$ new Pauli operators, whose number is unchanged by the following Clifford layers.
Thus, there are at most $\min ( (3n)^{\ell/d} , m) \cdot 2^{\mathcal{O}(\ell_t)}$ possible Pauli operators at the beginning of layer $t+1$.
Among these, we only count those of weight $\ell_{t+1}$.
At the next non-Clifford layer, each of these Pauli operators can split to at most $2^{\mathcal{O}(\ell_{t+1})}$ new Pauli operators, and so on. 
Iterating this argument, our upper bound on the number of Pauli paths is exactly the same as before [Eq.~(\ref{eq:number_paulip})], which leads to the same runtime as in Theorem~\ref{thm 1} of the main text.

\textbf{Gate-based noise:} We now turn to our algorithm for gate-based noise.
We incorporate the non-Clifford noise model into our algorithm by modifying our definition of the Pauli weight.
In our original algorithm, the Pauli weight served as a lower bound on the number of \emph{two-qubit gates} required to grow our initial operator into a Pauli operator of interest.
Since each two-qubit gate damped the operator amplitude by a factor $e^{-\gamma}$ due to noise, we formed an accurate classical approximation of our time-evolved operator by keeping track only of low-weight Pauli operators.
Applying this intuition to the non-Clifford noise model, we should introduce a modified weight that lower bounds the number of \emph{non-Clifford gates} required to transform our initial operator into some Pauli of interest.
This modification will depend on the circuit architecture (i.e.~the placement of the Clifford and non-Clifford gates) as well as the choice of Clifford gates.

To write this down explicitly, for each observable $O$, circuit $\mathcal{C}$, and circuit layer $t$, we define the \emph{non-Clifford distance} $w_{\text{nc}}[P;\mathcal{C},t]$ as the minimal number of non-Clifford gates in any Pauli path from $Q$ at layer zero to $P$ at layer $t$, where $Q$ is any Pauli operator with a non-zero coefficient in $O$. 
For example, at layer zero, the set of Pauli operators in $O$ have non-Clifford distance zero and all other Pauli operators have non-Clifford distance $\infty$.
At later layers, any Pauli operator that can be reached from $Q$ in $O$ via solely Clifford operations also has non-Clifford distance zero, at the layers at which it can be reached.
We emphasize that the definition of the non-Clifford distances changes as we move through the circuit; for example, a Pauli operator $Q$ in $O$ may have distance zero at time zero but distance $>0$ at time one, depending on the Clifford gates between time zero and one.

From this definition, we can also define a non-Clifford distance \emph{distribution}, $P^{(g)}_{\text{nc}}(w)$, and cumulative non-Clifford distance distribution, $Q^{(g)}_{\text{nc}}(w)$, for the time-evolved operator $O$.
These measure the norm of $O^{(g)}$ on Pauli operators with distance $w$ at circuit gate $g$, and are analogous to the weight distribution and cumulative weight distribution defined previously.
(One can think of this as working in a ``rotating frame'' determined by the action of the Clifford gates.)
From the definition of the non-Clifford distance, the distribution $P^{(g)}_{\text{nc}}(w)$ begins as a delta function at zero.
The distributions do not change under the action of any Clifford gate, and shift by at most one under each non-Clifford gate.

With these definitions, one can easily repeat our proof of Lemma~\ref{lemma: operator norm noise app} with non-Clifford gate-based noise instead of two-qubit gate-based noise, and Clifford distance instead of Pauli weight.
The only change is the factor of $e^{-4\gamma}$ is replaced with $e^{-2\gamma}$ throughout the proof, where the replacement arises because an operator can increase in non-Clifford distance despite only having support on one qubit in the gate and not two.
The end result is that at any layer $t$, the cumulative non-Clifford distance distribution is less than $Q^{(t)}_{\text{nc}}(w) \leq e^{-2\gamma(w+1)} \lVert O \rVert_F^2$ for all $w$.
If we form a low-distance approximation,
\begin{equation}
    \tilde{O}^{(t)}_{\text{nc}} = \sum_{P : w_{\text{nc}}[P;\mathcal{C},t] \leq \ell} \tilde{c}_P^{(t)} P,
\end{equation}
at each layer $t$,
then we similarly have  $\sum_{t=0}^d \tilde{Q}^{(t)}_{\text{nc}}(\ell) \leq e^{-2\gamma(\ell+1)} \lVert O \rVert_F^2$.

It remains only to bound the runtime of our classical approximation.
This is determined by the number of low-non-Clifford-distance Pauli operators that we must keep track of.
We begin with the operator $O$ acted on by the read-out noise channel.
Due to the read-out noise, we can truncate all Pauli operators in $O$ with weight above $\ell$.
From this starting point, each non-Clifford gate can increase the number of accessible Paulis by at most a constant factor $C$: for $T$ gates by a factor of 2, for general two-qubit non-Clifford gates by a factor of 15.
There are ${ n_{\text{nc}} \choose \ell }$ sets of $\ell$ non-Clifford gates.
Thus, there are at most $m { n_{\text{nc}} \choose \ell } C^\ell \leq m n_{\text{nc}}^\ell C^\ell / \ell! = \mathcal{O}(m n_{\text{nc}}^\ell)$ possible Pauli operators with non-Clifford distance $\leq \ell$ at any circuit layer, where $m$ is the number of Pauli operators with non-zero coefficients in $O$ and weight $\leq \ell$. 
The coefficients can be updated one-by-one in time $\mathcal{O}(m^2 n_{\text{nc}}^{2\ell})$.
Repeating $d$ times for each circuit layer, we arrive at the scaling in Eq.~(\ref{eq: non-clifford gate}).
\qed

\section{Extension to random non-unital noise} \label{app: non-unital}

We now consider the extension of our classical algorithms to quantum circuits with non-unital noise.
Unital noise corresponds to any noise process that maps the maximally mixed state to itself, a primary example being depolarizing noise.
Non-unital noise refers to noise processes that do not preserve the maximally mixed state, such as spontaneous emission noise (also referred to as amplitude damping).

Interestingly, the arguments in Ref.~\cite{ben2013quantum} prevent any immediate extension of our results to non-unital noise.
For any local non-unital noise model, Ref.~\cite{ben2013quantum} provides a scheme to perform fault-tolerant quantum computation for exponentially long times.
Their scheme assumes that the circuit begins in the zero state, in contrast to the input state ensembles that we consider.
However, for noise models that have a product state fixed point, such as spontaneous emission, this can easily be circumvented.
One simply needs to wait for some time $\mathcal{O}(\gamma^{-1} \log(n/\varepsilon))$ before applying the circuit in Ref.~\cite{ben2013quantum}, since after this time any input state will have become close to the fixed point.

While these arguments exclude classical algorithms for arbitrary quantum circuits with non-unital noise, recent work has shown that this may not be the case for restricted classes of circuits~\cite{mele2024noise}.
In particular, Ref.~\cite{mele2024noise} introduces a classical algorithm for expectation values in \emph{random circuits} with non-unital noise.
By taking each gate in the circuit to be independently random, one explicitly excludes the scheme in Ref.~\cite{ben2013quantum} which opens the door to an efficient classical simulation.
The algorithm in Ref.~\cite{mele2024noise}
runs in time $\exp(\mathcal{O}((\gamma^{-1} \log (\varepsilon^{-1}) )^D))$ for local observables in geometrically-local circuits in dimension $D$.
This is polynomial in $n$ for $D=1$, and quasi-polynomial for larger $D$.
For non-local observables, the runtime is instead exponential in $n$, and for circuits without geometric locality, it is exponential in $\varepsilon^{-1}$.

Our classical algorithms improve upon this performance in two ways.
First, we show that, in fact, a much milder form of randomness is sufficient to achieve an efficient classical simulation.
We elaborate on this in the following paragraph.
Second, we show that, under this mild randomness, our classical algorithms achieve runtimes matching those in Theorems~1,~2.
Namely, our algorithm runs in \emph{polynomial-time} for \emph{any} circuit architecture for large classes of observables, and quasi-polynomial time otherwise.

To elaborate on the first point, we consider quantum circuits with \emph{arbitrary}  gates, and require only that each non-unital noise channel occurs in a \emph{random direction}.
Considering spontaneous emission specifically, our requirement translates to choosing each local spontaneous emission channel to emit from either the 1 to 0 state,
\begin{equation}
    \mathcal{A}_{+} \{ \rho \} = \rho + \gamma_s \bra{ 1 } \rho \ket{ 1  } \dyad{0} - \frac{\gamma_s}{2} \dyad{1} \rho -\frac{\gamma_s}{2} \rho \dyad{1},
\end{equation}
or the 0 to 1 state,
\begin{equation}
    \mathcal{A}_- \{ \rho \} = \rho + \gamma_s \bra{ 0 } \rho \ket{ 0  } \dyad{1} - \frac{\gamma_s}{2} \dyad{0} \rho -\frac{\gamma_s}{2} \rho \dyad{0},
\end{equation}
with equal probability.
This randomization is performed implicitly in random circuits, via the random gates before and after each noise channel.
We emphasize that we are considering simulating a specific instance of the random non-unital noise (with high success probability over the random instances), and not a mixture of the instances.
Finally, as with our previous results, we consider the performance of our algorithm over an ensemble of input states.

This mild form of randomness neatly avoids the no-go results of Ref.~\cite{ben2013quantum}, by ensuring that the gates in the circuit are decoupled from the structure of the non-unital noise.
Thus, even though non-unital noise is present, the circuit cannot ``take advantage'' of it.
Within this setting, we show that our classical algorithms in Theorems~\ref{thm 1},~\ref{thm 2} of the main text apply equally well to circuits with non-unital noise.
\vspace{2mm}
\begin{extension} \label{ext: non-unital}
{\emph{(Classical algorithm for random non-unital noise)}}
Consider the settings of Theorems~\ref{thm 1},~\ref{thm 2} of the main text, but with each local depolarizing channel $\mathcal{D}_i$ replaced by spontaneous emission $\mathcal{A}_{i,\pm}$ in independently random directions with rate $\gamma_s \leq 4/7$.
Replace each error metric with its root-mean-square value over the random direction.
Then all of the stated results continue to hold, with $\gamma = \log(1/(1-\gamma_s/2)) \approx \gamma_s/2$ and the factor of $d$ in Eq.~(3) of the main text replaced by $1$.

\vspace{-3mm}
If one considers instead a random quantum circuit with uniform spontaneous emission noise, for any circuit architecture, then the runtime of our classical algorithm improves to
\begin{align}
        \emph{poly}(n) \cdot (1/\varepsilon)^{\mathcal{O}(1/\gamma)}, & \hspace{1cm} \text{if $O$ is a sum of polynomially many Paulis} \label{eq: thm1 uniform nonunital} \\
    n^{\frac{1}{\gamma} \log(1/\varepsilon)} \cdot (1/\varepsilon)^{\mathcal{O}(1/\gamma)}, & \hspace{1cm} \text{for any $O$}.
\end{align} 
\end{extension}
\noindent We make a few additional remarks.
First, for simplicity of analysis, we restrict attention to spontaneous emission noise  as the prototypical example of non-unital noise.
We expect our results to generalize straightforwardly to any strictly contractive non-unital noise channel~\cite{raginsky2002strictly}.
Second, the restriction to noise rates below a large threshold, $\gamma_s \leq 4/7$, is also chosen as a convenient simplification for our proof.
We do not expect larger noise rates to lead to harder classical simulations.
Finally, as a stepping stone in our proof, we prove a general lemma (Lemma~\ref{lemma: non-unital depth}) showing that expectation values in circuits with random non-unital noise depend only on the final $\mathcal{O}(\gamma^{-1}\log(n/\varepsilon))$ layers of the circuit for most input states.
This extends a result of Ref.~\cite{mele2024noise}, and follows extremely quickly after we introduce the general framework of our proof.

\emph{Proof of Extension~\ref{ext: non-unital}}---The strategy of our proof is to show that, while non-unital noise can increase the Frobenius norm of Heisenberg time-evolved operators in certain instances, it decreases the norm \emph{on average} over the random directions of the noise channel.
Thus, typical instances of non-unital noise share key properties in common with depolarizing noise.

To show this, let us begin by writing down the action of the spontaneous emission channel in the Heisenberg picture,
\begin{equation}
\begin{split}
    \mathcal{A}^{\dagger}_\pm \{ \mathbbm{1} \} & = \mathbbm{1} \\
    \mathcal{A}^{\dagger}_\pm \{ X \} & = (1-\gamma_s/2) X \\
    \mathcal{A}^{\dagger}_\pm \{ Y \} & = (1-\gamma_s/2) Y \\
    \mathcal{A}^{\dagger}_\pm \{ Z \} & = (1-\gamma_s) Z \pm \gamma_s \mathbbm{1}. \\
\end{split}
\end{equation}
The key idea we use is to decompose the spontaneous emission channel into the product,
\begin{equation}
    \mathcal{A}_\pm^\dagger = \tilde{\mathcal{A}}_\pm^\dagger \circ \mathcal{D},
\end{equation}
where the super-operator $\tilde{\mathcal{A}}$ is defined via
\begin{equation}
\begin{split}
    \tilde{\mathcal{A}}^{\dagger}_\pm \{ \mathbbm{1} \} & = \mathbbm{1} \\
    \tilde{\mathcal{A}}^{\dagger}_\pm \{ X \} & = X \\
    \tilde{\mathcal{A}}^{\dagger}_\pm \{ Y \} & = Y \\
    \tilde{\mathcal{A}}^{\dagger}_\pm \{ Z \} & = \frac{1-\gamma_s}{1-\gamma_s/2} Z \pm \frac{\gamma_s}{1-\gamma_s/2} \mathbbm{1}, \\
\end{split}
\end{equation}
and $\mathcal{D}$ is the depolarizing channel of strength $\gamma = -\log(1-\gamma_s/2) \approx \gamma_s/2$.

We will now show that, in expectation,  $\tilde{\mathcal{A}}_\pm^\dagger$ does not increase the Frobenius norm of any operator.
Suppose that $\tilde{\mathcal{A}}_\pm^\dagger$ acts on one qubit of a larger system.
Let us decompose an arbitrary operator $O$ in terms of its support on that qubit, $O = \mathbbm{1} \otimes O_\mathbbm{1} + X \otimes O_X + Y \otimes O_Y + Z \otimes O_Z$.
Considering the mean-square Frobenius norm over the two possible directions of the noise channel, we have
\begin{equation} \label{eq: expected Frobenius tilde A}
\begin{split}
    \mathbbm{E}_{\pm} \left[ \lVert \tilde{\mathcal{A}}_\pm^\dagger \{ O \} \rVert_F^2 \right] & = \lVert O_\mathbbm{1} \rVert_F^2 + \lVert O_X \rVert_F^2 + \lVert O_Y \rVert_F^2 + \left( \frac{(1-\gamma_s)^2}{(1-\gamma_s/2)^2} + \frac{\gamma_s^2}{(1-\gamma_s/2)^2} \right) \lVert O_Z \rVert_F^2 \\
    & \leq \lVert O \rVert_F^2 + \left( \frac{(1-\gamma_s)^2}{(1-\gamma_s/2)^2} + \frac{\gamma_s^2}{(1-\gamma_s/2)^2} - 1 \right) \lVert O_Z \rVert_F^2 \\
    & = \lVert O \rVert_F^2 - \gamma_s \left( \frac{1 - 7 \gamma_s / 4 }{(1-\gamma_s/2)^2} \right) \lVert O_Z \rVert_F^2 \\
    & \leq \lVert O \rVert_F^2,
\end{split}
\end{equation}
where the final inequality holds if $\gamma_s \leq 4/7$.
In the first line, we use that the cross terms between $Z$ and $\mathbbm{1}$ in $\tilde{\mathcal{A}}_\pm^\dagger\{ Z \}$ vanish in expectation.

To connect this to the accuracy of our classical algorithms, let us absorb each super-operator $\tilde{\mathcal{A}}_\pm$ into its adjacent unitary $U_t$, leaving only the depolarizing channels $\mathcal{D}$ as the noise channels.
We then analyze each setting in Extension~\ref{ext: non-unital} as follows.

\textbf{Gate-based noise: }For our Algorithm~2 for gate-based noise, the only properties of the unitaries $U_t$ that we used in our proofs were that: (i) each gate in $U_t$ cannot increase the weight of an operator by more than 1, and (ii) each gate cannot increase the Frobenius norm of an operator.
Since the super-operators $\tilde{\mathcal{A}}_\pm$ do not increase the operator weight, we are clearly allowed to absorb them into the $U_t$ with respect to condition (i).
In regards to condition (ii), by viewing our previous proof, one sees that the mean-square error over the input state ensemble depends only the Frobenius norms of the truncated operators.
Therefore, when taking the mean-square error over the random non-unital noise in addition, the error will only depend on the mean-square Frobenius norm over the random noise channels.
We have just shown that the super-operators $\tilde{\mathcal{A}}_\pm$ do not increase the mean-square Frobenius norm.
Thus, we are free to absorb the super-operators $\tilde{\mathcal{A}}_\pm$ into the unitaries $U_t$.
Our classical algorithm thus succeeds with runtime as in Theorem~\ref{thm 2} of the main text, with $\gamma = -\log(1-\gamma_s/2) \approx \gamma_s/2$. 

\textbf{Uniform noise: }For our Algorithm~1 for uniform noise, we require three small additional modifications.
The first two modifications arise because the presence of non-unital noise changes our counting of the number of valid Pauli paths, since non-unital noise can transform non-identity Pauli operators to identity Pauli operators (but not vice versa) in the Heisenberg picture.
This requires us to take Pauli paths with sequences of weights $(\ell_0,\ldots,\ell_t,0,\ldots,0)$ into consideration, where $\ell_0+\ldots+\ell_t \leq \ell$ and $t > 0$.
As our first modification, when bounding the error in Algorithm~1, these new Pauli paths lead to an increased prefactor of $\sum_{t=0}^d {\ell \choose t}$ instead of ${ \ell \choose d}$.
To derive this, note that in the first sum of Eq.~(\ref{eq:O-OF<l1l2}) we must now sum from $w_d = 0$ to $\ell-d+1$ instead of $w_d=1$ to $\ell-d+1$.
If $w_d > 0$, the entire bound proceeds as is and contains ${ \ell \choose d}$ individual truncations.
If $w_d=0$, then we are left with a bound resembling our original quantity of interest, but now with $d$ layers instead of $d+1$.
One can then apply the same argument to $w_{d-1}$ and so on.
For instance, the Pauli paths with $w_d=0$ and $w_{d-1}>0$ can be grouped into ${ \ell \choose d-1}$ individual truncations, and the Pauli paths with $w_d = w_{d-1}=0$ and $w_{d-2}>0$ can be grouped into ${ \ell \choose d-2}$ individual truncations, and so on.
In the end, this increased prefactor does not change the stated runtime of our algorithm, since when determining the required value of $\ell$ we used the inequality ${\ell \choose d} \leq (e \ell/d)^d$ [Eq.~(\ref{eq: binomial bound})].
This inequality is in fact somewhat weak, and is also obeyed by the partial sum of binomial coefficients, $\sum_{t=0}^d {\ell \choose t} \leq (e \ell/d)^d$.

Our second modification consists of a slight increase in the algorithm runtime due to the need to keep track of the new Pauli paths.
When $O$ is a sum of polynomially many $m$ Pauli operators, non-unital noise increases the number of Pauli paths by at most a factor of $d$, corresponding to the extra sum over $t =1,\ldots,d$.
For more general $O$, the sum over $t$ (with $\ell$ fixed) replaces the factor $n^{\ell/d}$ with $\sum_{t=1}^d n^{\ell/t} \leq d \cdot n^{\ell}$.

Finally, we must slightly modify our argument  (Lemma~\ref{lemma: uniform depth}) for restricting attention to circuits of depth $d < \gamma^{-1} \log(1/\varepsilon)$.
To do so, we provide a simple proof that observables in any circuit with random non-unital noise depend only on the final $\mathcal{O}(\gamma^{-1}\log(1/\varepsilon))$ layers of the circuit, for most input states.
This was proven recently for random quantum circuits using other techniques~\cite{mele2024noise}.
\vspace{2mm}
\begin{lemma} \label{lemma: non-unital depth}
{\emph{(Extension of Lemma~\ref{lemma: uniform depth} to random non-unital noise)}}
    Consider any observable $O$ and any circuit $\mathcal{C}$ with uniform noise, where each noise channel corresponds to spontaneous emission in a random direction with rate $\gamma_s \leq 4/7$.
    The root-mean-square Frobenius norm of the non-identity component of $\mathcal{C}^\dagger \{ O \}$ is upper bounded by $e^{-\gamma(d+1)} \lVert O \rVert_F$, where $d$ is the circuit depth and $\gamma = \log(1/(1-\gamma_s/2)) \approx \gamma_s/2$.
\end{lemma}
\vspace{2mm}
\noindent \emph{Proof}---We decompose each noise channel as $\mathcal{A}_\pm^\dagger = \tilde{\mathcal{A}}_\pm^\dagger \circ \mathcal{D}$, and apply Eq.~(\ref{eq: expected Frobenius tilde A}) to the proof of Lemma~\ref{lemma: uniform depth}. \qed
\vspace{2mm}

\noindent The lemma implies that we can approximate $\mathcal{C}^\dagger \{ O \}$ for depths $d \geq d^* = \gamma^{-1} \log(1/\varepsilon)$ by its identity component, $\text{tr}( O^{(d^*)} )/2^n \cdot \mathbbm{1}$,  at layer $d^*$.
This follows because the non-identity components have small root-mean-square Frobenius norm, and the layers after $d^*$ map the identity operator to itself in the Heisenberg picture.
From Lemma~\ref{lemma: average error app}, this approximation suffices to compute expectation values to within root-mean-square error $\varepsilon \cdot \sqrt{c} \cdot \lVert O \rVert_F$ over any low-average ensemble of input states with purity $c$. 

\textbf{Random quantum circuits with uniform noise: }Finally, we demonstrate the improved runtime of our Algorithm~1 when applied to random quantum circuits.
Compared to the general quantum circuits that we consider, the main simplification that occurs for random quantum circuits is that our error bound for Algorithm~1 improves to $e^{-\gamma(\ell+1)} \lVert O \rVert_F$ instead of $e^{-\gamma(\ell+1)} \sqrt{\sum_{t=0}^d {\ell \choose t}} \lVert O \rVert_F$.
This follows because different Pauli paths, on average, do not coherently interfere in random quantum circuits (we refer to Ref.~\cite{aharonov2023polynomial} for a comprehensive discussion).
This improved error bound allows us to choose a much smaller threshold $\ell = \gamma^{-1} \log(1/\varepsilon)$.
This gives a total number of low-weight Pauli paths 
\begin{equation}
    \min(d \cdot n^{\ell}, d \cdot m) \cdot 2^{\mathcal{O}(\ell)} = d \cdot \min( n^{\frac{1}{\gamma} \log(1/\varepsilon)} , m ) \cdot (1/\varepsilon)^{\mathcal{O}(1/\gamma)}.
\end{equation}
The factor of $d$ can be absorbed into the factor $(1/\varepsilon)^{\mathcal{O}(1/\gamma)}$, giving our stated runtime. \qed

\section{Implications for quantum error mitigation} \label{app: error mit}

In this section, we prove Corollary~\ref{cor: error mitigation} of the main text.
We note that the corollary captures nearly all error mitigation strategies of interest, including zero-noise extrapolation~\cite{li2017efficient,temme2017error}, probabilistic error cancellation~\cite{temme2017error},  and Clifford data regression~\cite{czarnik2021error}.
One notable exception are \emph{virtual distillation} protocols~\cite{huggins2021virtual,o2021error}, which evade the settings of our work by performing experiments involving multiple copies of each input state.
Another (partial) exception are noise \emph{suppression} strategies, such as dynamical decoupling~\cite{viola1999dynamical,khodjasteh2005fault}, which serve to decrease the effective noise rate $\gamma$ (and thus, increase the potential classical complexity), by addressing the microscopic, unitary source of the noise.
We remark that, in the corollary, the root-mean-square error for an error mitigation strategy is necessarily taken over both the ensemble of input states, as well as the probabilistic measurement outcomes of the noisy quantum experiments themselves.

\emph{Proof of Corollary~\ref{cor: error mitigation}}---We denote the set of noisy quantum circuits as $\{ \mathcal{C}' \}$ and the number of circuits as $M = | \{ \mathcal{C}' \} |$.
To prove the corollary, we adapt our classical algorithms to allow sampling from the measurement outcomes of $O$  in each of the $M$ noisy circuits $\mathcal{C}'$.
Without loss of generality, we can assume that the measurements occur in the eigenbasis of $O$, so that each outcome is some eigenvalue $\lambda$.
The outcomes are received  with probabilities $p_{\mathcal{C}'\{\rho\}}(\lambda) = \tr( P_\lambda \mathcal{C}' \{ \rho \})$, where $P_\lambda$ is the projector onto the $\lambda$ eigenspace of $O$.
Thus, to classically reproduce the result of the error mitigation strategy, we need to sample once from $p_{\mathcal{C}'\{\rho\}}(\lambda)$ for each $\mathcal{C}'$.
We will show that this can be done with minimal overhead for Pauli observables, since each $\lambda$ takes only two possible eigenvalues $\pm1$.

To sample from each individual $p_{\mathcal{C}'\{\rho\}}(\lambda)$, we first classically compute the expectation value $\tr(O \mathcal{C}' \{ \rho \})$ to within error $\varepsilon'$.
Denote our classical approximation as $a_{\mathcal{C}'}$.
From this, we approximate the true outcome distribution $p_{\mathcal{C}'\{\rho\}}(\pm 1)$ via $q_{\mathcal{C}'\{\rho\}}(\pm 1) = (1\pm a_{\mathcal{C}'})/2$. 
The approximation has total variational distance $\sum_\pm | p_{\mathcal{C}'\{\rho\}}(\pm 1) - q_{\mathcal{C}'\{\rho\}}(\pm 1)| \leq \varepsilon'$.
We can sample from $q_{\mathcal{C}'\{\rho\}}(\pm 1)$ efficiently since it takes only two values.

We now analyze how the individual sampling errors propagate when we sample from all $M$ noisy circuits $\{\mathcal{C}'\}$ in succession.
The ideal outcome distribution is the tensor product $p(\textbf{r}) = \bigotimes_{m=1}^M p_{\mathcal{C}_r'\{\rho\}}(r_i)$. Here, $\textbf{r} = (r_1,\ldots,r_M)$ for $r_m \in \{-1,1\}$ is the sequence of measurement outcomes, and $p(\textbf{r})$ is the probability for a set of $M$ experiments to receive the sequence.
Our approximation is the tensor product $q(\textbf{r}) = \bigotimes_{m=1}^M q_{\mathcal{C}_r'\{\rho\}}(r_i)$.
We will bound the total variational distance between $p(\textbf{r})$ and $q(\textbf{r})$.
Denote the total variational distance as $\text{TVD}(p,q) = \sum_{\textbf{r}} | p(\textbf{r}) - q(\textbf{r})|$.
We use the fact that the total variational distance is at most additive under the tensor product,
\begin{equation}
    \text{TVD}(p_1 \otimes p_2,q_1 \otimes q_2) \leq \text{TVD}(p_1,q_1) + \text{TVD}(p_2,q_2).
\end{equation}
Applied $M$ times, this gives $\text{TVD}(p,q) \leq M \varepsilon'$.

To conclude our proof, we note that any error mitigation strategy must assign some estimate $a(\textbf{r})$ of $\tr( O \mathcal{C} \{ \rho \})$, for every possible sequence $\textbf{r}$ of measurement outcomes.
Without loss of generality, we can assume that $|a(\textbf{r})| \leq 1$, since $\lVert O \rVert_\infty = 1$ for any Pauli operator.
Now, we can utilize the same assignment $a(\textbf{r})$ for our classical algorithm.
To bound the mean-square error in our classical estimate, we use
\begin{equation}
\begin{split}
    \sum_{\textbf{r}} q(\textbf{r}) \left( O(\textbf{r}) - \tr( O \mathcal{C} \{ \rho \} ) \right)^2 & = \sum_{\textbf{r}} p(\textbf{r}) \left( O(\textbf{r}) - \tr( O \mathcal{C} \{ \rho \} ) \right)^2 + \sum_{\textbf{r}} (p(\textbf{r})- q(\textbf{r})) \cdot \left( O(\textbf{r}) - \tr( O \mathcal{C} \{ \rho \} ) \right)^2 \\
    & \leq \sum_{\textbf{r}} p(\textbf{r}) \left( O(\textbf{r}) - \tr( O \mathcal{C} \{ \rho \} ) \right)^2 + 4 \text{TVD}(p,q) \\
    & \leq \sum_{\textbf{r}} p(\textbf{r}) \left( O(\textbf{r}) - \tr( O \mathcal{C} \{ \rho \} ) \right)^2 + 4 M \varepsilon'. \\
\end{split}
\end{equation}
The first term is the root-mean-square error in the error mitigation strategy.
Thus, if error mitigation can compute the observable to error $\varepsilon/2$,  our classical algorithm can do so to  root-mean-square error $\varepsilon$ by taking $\varepsilon' = \varepsilon/8M$.
By assumption, $\varepsilon^{-1}$ and $M$ are both polynomial in $n$.
Hence, $\varepsilon'^{-1}$ is also polynomial in $n$. 
Thus, by Theorem~1, the computation of $\tr(O \mathcal{C}' \{ \rho \})$ above can be done in polynomial time, and so the ideal expectation value can also be computed in polynomial time.
\qed

\section{Test for the  hardness of a given quantum circuit} \label{app: cor 3}

Let us now turn to Corollary~\ref{cor: sensitivity}. 
In general, classical simulation of noiseless quantum circuits will require exponential resources.
However, this is not guaranteed for every such circuit.
Motivated by modern quantum experiments, in the main text we proposed Corollary~\ref{cor: sensitivity} as a simple test for whether a given quantum circuit has the potential to achieve a quantum advantage.
We remark that the converse of the corollary clearly does not hold. Namely, it is easy to design experiments that are highly sensitive to noise by considering, for example, a many-body observable~\cite{kim2023evidence} or a Loschmidt echo~\cite{goussev2012loschmidt}, yet such experiments are not guaranteed to be hard classically~\cite{rudolph2023classical,schuster2023operator}.

\emph{Proof of Corollary~\ref{cor: sensitivity}}---We denote the noiseless quantum circuit as $\mathcal{C}$, and the same circuit with  noise as $\mathcal{C}_\gamma$ for a noise rate $\gamma$.
Let $O^{(d)} = \mathcal{C}^\dagger\{ O \}$ denote the Heisenberg time-evolved observable under the noiseless circuit, $O_\gamma^{(d)} = \mathcal{C}^\dagger_\gamma\{ O \}$ under the noisy circuit, and $\tilde{O}^{(d)}$ be the approximation obtained by applying our classical algorithm to $\mathcal{C}_\gamma$.
For each input state $\rho$, we have
\begin{equation}
	\big| \tr \big( \rho O^{(d)} \big)  - \tr \big( \rho \tilde{O}^{(d)} \big)  \big| \leq \big| \tr \big( \rho O^{(d)} \big) - \tr \big( \rho O_\gamma^{(d)} \big)  \big|  + \big| \tr \big( \rho O_\gamma^{(d)} \big)  - \tr \big( \rho \tilde{O}^{(d)} \big) \big|.
\end{equation}
Suppose that the noisy quantum experiment succeeds in estimating the noiseless expectation value to within root-mean-square error $\varepsilon$, for some noise rate $\gamma$.
This implies that the root-mean-square of the first term on the RHS is less than $\varepsilon$.
Meanwhile, the root-mean-square of the second term on the RHS is less than $\varepsilon$ if we take $\ell$ as in Theorem~\ref{thm 2} of the main text.
Thus, the classical algorithm in Theorem~2 can compute expectation values within root-mean-square error $4\varepsilon$ in time as stated in Theorem~2.
Imposing the lower bound $\chi(n)$ on the runtime of the classical algorithm, setting $d,\varepsilon^{-1} = \text{poly}(n)$, and solving for the noise rate $\gamma$ gives our desired result.\qed

\section{Noisy quantum circuits on any highly mixed input state} \label{app: cor 1}

As mentioned in the main text, our results in Theorems~1,~2 immediately extend to circuits on any highly mixed input state.
This follows because the ensemble composed of a single state $\rho$ is low-average whenever $\lVert \rho \rVert_\infty = c/2^n$ with $c = \mathcal{O}(1)$.
In fact, we can prove a slightly stronger statement, in which this condition on the spectral norm is replaced by a condition on the Renyi-2 entropy, $S^{(2)}(\rho) = \log_2( 1/\lVert \rho \rVert_2^2 )$.
\vspace{2mm}
\begin{corollary} \label{cor: mostly mixed}
{\emph{(Classical algorithms for noisy quantum circuits on highly mixed states)}}
{Consider a noisy circuit $\mathcal{C}$, and a state $\rho$ with Renyi-2 entropy $S^{(2)}(\rho) = n - \log_2(c)$.
Assume the Pauli coefficients of $\rho$ can be efficiently computed.
There exists a classical algorithm to compute the state $\mathcal{C}\{\rho\}$ to within trace norm error $\varepsilon \cdot \sqrt{c}$ in time as in Theorems~1,~2.}
\end{corollary}
\vspace{2mm}

\emph{Proof of Corollary~\ref{cor: mostly mixed}}---The corollary follows by applying our Algorithms~1,~2 to the density matrix $\rho$ instead of an observable.
This produces a classical approximation $\tilde{\rho}$ of the density matrix.
The approximation has unit trace but is not guaranteed to be positive.
We quantify the error, $\delta \rho = \tilde{\rho} - \mathcal{C} \{ \rho \}$, in the approximation via the trace norm $\lVert \delta \rho \rVert_1$.
We have
\begin{equation}
	\lVert \delta \rho \rVert_1 \leq \sqrt{2^n} \cdot \lVert \delta \rho \rVert_2 \leq \varepsilon \cdot \sqrt{2^n} \cdot \lVert \rho \rVert_2 = \varepsilon \cdot \sqrt{c},
\end{equation}
where the second inequality follows from the proofs of Theorems~1,~2, and the final inequality uses the definition of the Renyi-2 entropy. \qed

\section{Noisy quantum circuits with spatial disorder} \label{app: cor 2}

We can also exchange the ensemble of input states in Theorems~1,~2 with an ensemble of \emph{quantum circuits} applied to any fixed input state.
Intuitively, this follows by absorbing the preparation of the input state ensemble into the circuit itself. 
We illustrate this  with a simple example.
Consider estimating the expectation value of a Pauli operator $Q$ over the ensemble of computational basis states.
We can write each state $\ket{\textbf{s}}$ as a Pauli operator applied to the zero state, $\ket{\textbf{s}} = X_{\textbf{s}} \ket{0^n}$, with $X_\textbf{s} = \bigotimes_{i=0}^n X_i^{s_i}$.
The expectation values are
\begin{equation} \label{eq: state to circuit and obsv}
	\tr( \mathcal{C} \{ \dyad{{\textbf{s}}} \} Q ) = \tr( \mathcal{C} \{ X_\textbf{s} \dyad{0^n} X_\textbf{s}^\dagger \} \cdot Q ) = \tr( \mathcal{C}^{X_\textbf{s}} \{  \dyad{0^n}  \} \cdot X_\textbf{s}^\dagger Q X_\textbf{s} ) = (-1)^{a[X_\textbf{s},Q]} \cdot \tr( \mathcal{C}^{X_\textbf{s}} \{  \dyad{0^n}  \} \cdot Q  ),
\end{equation}
where $a[X_\textbf{s},Q]$ is 0 if $X_\textbf{s}$ and $Q$ commute and 1 if they anti-commute.
Here, we define the conjugated circuit $\mathcal{C}^P\{ ( \cdot ) \} = P^\dagger \mathcal{C} \{ P ( \cdot ) P^\dagger \} P$, which replaces each unitary in the original circuit with its conjugation by $P$, $U_t \rightarrow P^\dagger U_t P$. 
We note that multiplying by an overall phase does not change the estimation error.
Thus, since our classical algorithm can estimate expectation values for the fixed circuit $\mathcal{C}$ applied to the  ensemble of computational basis states, it can also estimate expectation values for the ensemble of circuits $\{ \mathcal{C}^{X_\textbf{s}} \}$ applied to the fixed input state $\dyad{0^n}$.

We generalize this example in two ways.
First, we can extend beyond Pauli observables by decomposing an arbitrary observable in the Pauli basis, $O = \sum_Q c_Q Q$.
Since each Pauli can be estimated within error $\varepsilon$,  the operator $O$ can be estimated within error $\varepsilon \cdot \lVert O \rVert_{\text{Pauli},1}$, where
$
	\lVert O \rVert_{\text{Pauli},1} = \sum_Q | c_Q |
$
is the Pauli 1-norm.
For most observables of interest, the Pauli 1-norm is upper bounded by the Frobenius and spectral norms.
In particular, we have $\lVert O \rVert_{\text{Pauli,1}} \leq \sqrt{m}  \lVert O \rVert_F$ for observables that are sums of $m$ Pauli operators, and $\lVert O \rVert_{\text{Pauli},1} \leq C(d,k)  \lVert O \rVert_\infty$~\cite{huang2022learning} for $k$-local observables of degree $d$, where $C(d,k) = 3 \sqrt{d} \exp( \Theta( k \log k) )$.

Second, we can replace the specific circuit ensemble $\{ \mathcal{C}^{X_\textbf{s}} \}$ with any circuit ensemble that is invariant under conjugation by $X_\textbf{s}$, i.e.~$\{ \mathcal{C}^{X_\textbf{s}} \} = \{ \mathcal{C} \}$ for all $\textbf{s}$.
Since the single-qubit Pauli-$X$ operators generate the full set of $X_\textbf{s}$, this is equivalent to the ensemble being invariant under conjugation by single-qubit Pauli-$X$ operators.
Hence, we refer to this as \emph{spatial disorder} in the circuit ensemble.
In the case that the ensemble is invariant under \emph{any} single-qubit Pauli operator, we can extend this result to arbitrary input states.
This follows since $\{ P \rho P^\dagger | \mathcal{P}_n \}$ is a low-average ensemble with $c=1$ for any fixed $\rho$ when $P$ is drawn from the $n$-qubit Pauli group $\mathcal{P}_n$.
We capture both of these settings with the following definition.
\vspace{2mm}
\begin{definition} \label{def: spatial disorder}
{\emph{(Spatial disorder in quantum circuit ensembles)}}
{We say that an ensemble of quantum circuits $\mathcal{E}_\mathcal{C} = \{ \mathcal{C} \}$ has \emph{spatial disorder} if the ensemble is invariant under conjugation by any single-qubit Pauli operator $P$,
\begin{equation}
	\{ \mathcal{C}^P  \} = \{ \mathcal{C} \}.
\end{equation}
As a weaker condition, we say that an ensemble $\mathcal{E}_\mathcal{C}$ has \emph{spatial disorder on a quantum state} $\rho$ if the same condition holds for a subset $\mathcal{E}_P$ of Paulis such that $\frac{1}{|\mathcal{E}_P|} \sum_P P \rho P^\dagger = \mathbbm{1}/2^n$.
}
\end{definition}
\vspace{2mm}
\noindent Our definition encompasses numerous circuits of interest for the quantum simulation of disordered spin models, including multiple recent experiments~\cite{choi2017observation,randall2021many,mi2022time,harris2018phase,king2023quantum,king2024computational}.
For example, a circuit composed of unitaries of the form
\begin{equation}
	U_t  = \exp \bigg( i \sum_i g_i X_i \bigg) \cdot \exp \bigg(i \sum_{(i,j) \in U_t} J_{ij} Z_i Z_j \bigg),
\end{equation}
has spatial disorder if $g_i$, $J_{ij}$ are drawn at random from any distributions that are symmetric about zero.
We can also allow, for instance, the $g_i$ to be non-random if the input state is diagonal in the $y$ or $z$-basis, or the $J_{ij}$ to be non-random for the $x$ or $y$-basis.

Leveraging our Theorems~1,~2, we show that noisy quantum circuits with spatial disorder can be efficiently classically simulated.
\vspace{2mm}
\begin{corollary} \label{cor: spatial disorder app}
{\emph{(Classical algorithm for noisy quantum circuits with spatial disorder)}}
{Consider a state $\rho$, an observable $O$, and an ensemble $\mathcal{E}_\mathcal{C} = \{\mathcal{C}\}$ of noisy quantum circuits with spatial disorder on $\rho$.
Assume the Pauli coefficients of $\rho$ and $O$ can be efficiently computed.
There exists a classical algorithm to compute the expectation values $\tr( \mathcal{C} \{ \rho \} O)$ to within root-mean-square error $\varepsilon \cdot \lVert O \rVert_{\emph{Pauli},1}$ in time as in Theorems~\ref{thm 1},~\ref{thm 2} of the main text.}
\end{corollary}
\vspace{2mm}

\emph{Proof of Corollary~\ref{cor: spatial disorder app}}---We use the same classical algorithms as in Theorems~1,~2.
The truncations in our classical algorithms respect additivity, so that our approximation for $O$ under circuit $\mathcal{C}$
 is equal to a sum of the corresponding approximations for each Pauli operator,
 \begin{equation}
 \tilde{O}_\mathcal{C}^{(d)} = \sum_Q c_Q \cdot \tilde{Q}^{(d)}_\mathcal{C},
\end{equation}
 where the coefficients $c_Q$ are obtained from the Pauli decomposition of $O$ before time-evolution. 
 If we consider instead the conjugated quantum circuit $\mathcal{C}^P$ for some Pauli $P$, we have
 \begin{equation}
	\tilde{O}^{(d)}_{\mathcal{C},P} = \sum_Q (-1)^{a[P,Q]} \cdot c_Q \cdot P^\dagger \tilde{Q}_{\mathcal{C}}^{(d)} P.
\end{equation}
Thus, the difference between the exact and approximate operator is
\begin{equation}
	O^{(d)}_{\mathcal{C},P} - \tilde{O}^{(d)}_{\mathcal{C},P}  = \sum_Q (-1)^{a[P,Q]} \cdot c_Q \cdot P^\dagger \left( Q_\mathcal{C}^{(d)} - \tilde{Q}_\mathcal{C}^{(d)} \right) P.
\end{equation}

By assumption, the ensemble $\mathcal{E}_\mathcal{C}$ is invariant under conjugation by any Pauli operator, since this follows from it being invariant under local Pauli conjugations.
Thus, we can replace the average over $\mathcal{E}_\mathcal{C}$ with an identical average over the ensemble $\mathcal{E}_{\mathcal{C}^P} \equiv \{ \mathcal{C}^P | \mathcal{C} \in \mathcal{E}_\mathcal{C}, P \in \mathcal{P}_n \}$, where $\mathcal{P}_n$ is the set of all $n$-qubit Pauli operators.
The mean-square error over $\mathcal{E}_{\mathcal{C}^P}$ is
\begin{equation}
\begin{split}
	\frac{1}{4^n | \mathcal{E}_\mathcal{C} |}  \sum_{\mathcal{C},P} \tr(  \rho \cdot (O^{(d)}_{\mathcal{C},P} - \tilde{O}^{(d)}_{\mathcal{C},P}) )^2 & \leq  \frac{1}{4^n | \mathcal{E}_\mathcal{C} |}  \sum_{\mathcal{C},P} \left( \sum_Q | c_Q | \cdot \left| \tr( P \rho P^\dagger \cdot  (Q_\mathcal{C}^{(d)} - \tilde{Q}_\mathcal{C}^{(d)}) ) \right| \right)^2 \\
	& \leq \frac{1}{4^n | \mathcal{E}_\mathcal{C} |}  \sum_{\mathcal{C},P, Q,Q'} | c_Q | | c_{Q'} | \left| \tr( P \rho P^\dagger \cdot \delta Q_\mathcal{C} ) \right| \left| \tr( P \rho P^\dagger \cdot \delta Q_\mathcal{C}' ) \right| \\
	& \leq \frac{1}{| \mathcal{E}_\mathcal{C} |}  \sum_{\mathcal{C}, Q,Q'} | c_Q | | c_{Q'}|   \left(  \frac{1}{4^n} \sum_P \tr( P \rho P^\dagger \cdot \delta Q_\mathcal{C} )^2 \right)^{1/2} \left(  \frac{1}{4^n}  \sum_P \tr( P \rho P^\dagger \cdot \delta Q_\mathcal{C}' )^2 \right)^{1/2}, \\
\end{split}
\end{equation}
where the last line uses the Cauchy-Schwarz inequality, and we abbreviate $\delta Q_\mathcal{C} = Q_\mathcal{C}^{(d)} - \tilde{Q}_\mathcal{C}^{(d)}$.
Now, note that the terms within parentheses correspond to mean-square error of $Q$ and $Q'$ under the \emph{state ensemble} $\mathcal{E} = \{ P \rho P^\dagger | P \in \mathcal{P}_n \}$.
By assumption, this is a zero-average ensemble.
Thus, by the results of Theorems~1,~2, the term is upper bounded $ \varepsilon^2$.
We have
\begin{equation}
\begin{split}
	\frac{1}{4^n | \mathcal{E}_\mathcal{C} |}  \sum_{\mathcal{C},P} \tr(  \rho \cdot (O^{(d)}_{\mathcal{C},P} - \tilde{O}^{(d)}_{\mathcal{C},P}) )^2 & \leq \frac{1}{|\mathcal{E}_\mathcal{C}|} \sum_\mathcal{C} \sum_{Q,Q'} |c_Q| |c_Q'| \cdot \varepsilon^2 = \varepsilon^2 \bigg( \sum_Q | c_Q | \bigg)^2 = \varepsilon^2 \cdot \lVert O \rVert_{\text{Pauli},1}^2. \\
\end{split}
\end{equation}
as desired. \qed

\end{document}